\documentclass[journal]{IEEEtran}
\usepackage[utf8]{inputenc}

\ifCLASSINFOpdf
   \usepackage[pdftex]{graphicx}
\else
   \usepackage[dvips]{graphicx}
\fi
\usepackage{color}
\usepackage{xcolor}
\usepackage{amsmath,amsfonts}
\usepackage{commath}
\usepackage{mathtools}
\usepackage{gensymb}
\usepackage{tabularx,booktabs}
\usepackage{textcomp}
\usepackage{verbatim}
\usepackage{upgreek}
\usepackage{color}
\usepackage{soul}
\usepackage{cite}
\usepackage{bm}
\usepackage{threeparttable}
\usepackage[labelformat=simple]{subcaption}
\usepackage{outlines}
\usepackage{lipsum}
\usepackage{array}
\usepackage{tikz, pgfplots}
\usetikzlibrary{positioning, decorations.pathreplacing,calligraphy}
\usetikzlibrary{math}
\pgfplotsset{compat=1.18}
\usepackage[inline]{enumitem}

\usepackage{hyperref}
\hypersetup{
    colorlinks=true,
    linkcolor=black,
    filecolor=magenta,      
    urlcolor=teal,
    citecolor=black,
    pdftitle={Ultra-Low-Loss Fiber Bragg Grating Mode Scrambler Design Exploiting Propagation Constant Engineering},
    }
\usepackage{cleveref}

\newcommand\lenS{L_S}
\newcommand\Linter{L_{\mathrm{inter}}}

\newcommand{\E}[1]{\ensuremath{\mathbb{E}\left\{{#1} \right\}}}

\newcommand{\PC}{\ensuremath{\bm{\mathcal{P}}}}
\newcommand{\PCMP}{\ensuremath{\bm{\mathcal{P}}_\mathrm{MP}}}
\newcommand{\RMP}{\ensuremath{\bm{R}_\mathrm{MP}}}
\newcommand{\DPC}{\ensuremath{\mathbf{D}^{-1} \bm{\mathcal{P}}}}
\newcommand{\tautau}{\ensuremath{\bm{{\tau}}}}

\newcommand{\ttot}{\bm{\tau}_\textrm{tot}}

\newcommand\bvec{\boldsymbol{{\beta}}}

\newcommand\LP{\mathrm{LP}}
\newcommand\HE{\mathrm{HE}}
\newcommand\TE{\mathrm{TE}}
\newcommand\TM{\mathrm{TM}}
\definecolor{cardinal}{RGB}{45,45,45}
\definecolor{berkeley}{RGB}{0,50,200}
\definecolor{graydark}{RGB}{45,45,45}

\newcommand{\drawStepIndexAnalysis}[0]
{
    \tikzmath {
        \scale = 5;
        \nefflpA1 = 0.890594486812127 * \scale;
        \nefflpB1 = 0.724326744127913 * \scale;
        \nefflpC1 = 0.509701074979929 * \scale;
        \nefflpA2 = 0.440473230409359 * \scale;
        \nefflpD1 = 0.253471291179478 * \scale;
        \nefflpB2 = 0.130147296609290 * \scale;
        \dbscale = 8;
        \dboffset = -3;
        \delbeta1 = 0.326206380260724 * \dbscale + \dboffset;
        \delbeta2 = 0.556901932627356 * \dbscale + \dboffset;
        \delbeta3 = 0.421081452803370 * \dbscale + \dboffset;
        \delbeta4 = 0.366885510761890 * \dbscale + \dboffset;  
        \delbeta5 = 0.608839080459623 * \dbscale + \dboffset;
        \delbeta6 = 0.502705990585877 * \dbscale + \dboffset;
        \delbeta7 = 0.744659560283610 * \dbscale + \dboffset;
        \delbeta8 = 0.864179520176013 * \dbscale + \dboffset;
        \delbeta9 = 1.000000000000000 * \dbscale + \dboffset;
        \ncore    = 1 * \scale;
        \x0 = -8;
        \y0 = 0;
        \rcore = 2;
        \rclad = 4;
        \rend = 4.3;
        \yend = 1.1 * \scale;
        \yax = 0.3;
        \xp0 = \x0+7;
        \barlen = 3;
        \labelshift = 0.3;
        \redshift = 0.3;
        \minishift = 0.05;
        \redyshift = 0.5;
        \redyoffset = 0.3;
        \redyoffsetb = 0.4;
        \xdb0 = \xp0 + 6;
    }
    \useasboundingbox (-4.0,-1.2) rectangle (6.0,6.5);
    \draw[ultra thick, black, -stealth] (\x0,\y0-\yax) -- (\x0+\rend,\y0-\yax) node[midway, yshift=-0.3cm] {Radial Position $r$};
    \draw[ultra thick, black, -stealth] (\x0,\y0-\yax) -- (\x0,\y0+\yend) node[midway, xshift=-0.3cm, rotate=90] {Refractive Index $n_\mathrm{init}(r)$};
    \draw[ultra thick, black] (\x0,\y0 + \ncore) -- (\x0+\rcore, \y0 + \ncore) -- (\x0+\rcore, \y0) -- (\x0+\rclad, \y0);
    \node at (\x0/2 + \rend/2-4-0.5,\y0+\yend+0.5)    {\textbf{Transverse Index Profile}};
    \node at (\xp0 + \rend/2-1,\y0+\yend+0.5)     {\textbf{Propagation Constants}};
    \node at (\xdb0 + \rend/2-0.25,\y0+\yend+0.5) {\textbf{Propagation Constant Spacings}};
    \node at (\x0/2 + \rend/2-4,\y0-1)    {\textbf{(a)}};
    \node at (\xp0 + \rend/2-1,\y0-1)     {\textbf{(b)}};
    \node at (\xdb0 + \rend/2-0.25,\y0-1) {\textbf{(c)}};
    
    \node at (\x0+\rcore/2 + \rend/2,\y0+0.25) {\textcolor{black}{{$n_{\mathrm{clad}}$}}};
    \draw[ultra thick, black] (\xp0,\nefflpA1) node[anchor=east]{{$\mathrm{LP}_{01}$}} -- (\xp0+\barlen,\nefflpA1);
    \draw[ultra thick, black] (\xp0,\nefflpB1)  node[anchor=east]{{$\mathrm{LP}_{11}$}} -- (\xp0+\barlen,\nefflpB1);
    \draw[ultra thick, black] (\xp0,\nefflpC1) node[anchor=east]{{$\mathrm{LP}_{21}$}} -- (\xp0+\barlen,\nefflpC1);
    \draw[ultra thick, black] (\xp0,\nefflpA2) node[anchor=east]{{$\mathrm{LP}_{02}$}} -- (\xp0+\barlen,\nefflpA2);
    \draw[ultra thick, graydark] (\xp0,\nefflpD1) node[anchor=east]{{$\mathrm{LP}_{31}$}} -- (\xp0+\barlen,\nefflpD1);
    \draw[ultra thick, graydark] (\xp0,\nefflpB2) node[anchor=east]{{$\mathrm{LP}_{12}$}} -- (\xp0+\barlen,\nefflpB2);
    \draw[ultra thick, graydark] (\xp0,\y0) node[anchor=east]{{$k_{0}\mathrm{n}_{\mathrm{clad}}$}} -- (\xp0+\barlen,\y0);
    
    \draw[thick, stealth-stealth, berkeley] (\xp0+\barlen,\nefflpA1) -- (\xp0+\barlen,\nefflpB1);
    \draw[thick, stealth-stealth, berkeley] (\xp0+\barlen,\nefflpB1) -- (\xp0+\barlen,\nefflpC1);
    \draw[thick, stealth-stealth, berkeley] (\xp0+\barlen-\labelshift,\nefflpB1) -- (\xp0+\barlen-\labelshift,\nefflpA2);

    \draw[thick,-stealth, berkeley, dashed] (\xp0+\barlen-0*\labelshift+\minishift,\nefflpA1 / 2 + \nefflpB1 / 2) -- (\xp0+\barlen+\redshift,\nefflpA1 - 0 * \redyshift - \redyoffsetb)  node[anchor=west]{$\Delta\beta_{01,11}$};
    \draw[thick,-stealth, berkeley, dashed] (\xp0+\barlen-0*\labelshift+\minishift,\nefflpB1 / 2 + \nefflpC1 / 2) -- (\xp0+\barlen+\redshift,\nefflpA1 - 1 * \redyshift - \redyoffsetb)  node[anchor=west]{$\Delta\beta_{11,21}$};
    \draw[thick,-stealth, berkeley, dashed] (\xp0+\barlen-1*\labelshift+\minishift,\nefflpB1 / 2 + \nefflpA2 / 2) -- (\xp0+\barlen+\redshift,\nefflpA1 - 2 * \redyshift - \redyoffsetb)  node[anchor=west]{$\Delta\beta_{11,02}$};

    \draw[thick, stealth-stealth, cardinal] (\xp0+\barlen,\nefflpC1) -- (\xp0+\barlen,\nefflpD1);
    \draw[thick, -stealth, cardinal, dashed] (\xp0+\barlen+\minishift,\nefflpC1 / 2 + \nefflpD1 / 2 + 0.2) -- (\xp0+\barlen+\redshift,\nefflpA2+\redyshift - \redyoffset) node[anchor=west]{$\Delta\beta_{21,31}$};
    
    \draw[thick, stealth-stealth, cardinal] (\xp0+\barlen-2*\labelshift,\nefflpC1) -- (\xp0+\barlen-2*\labelshift,\nefflpB2);
    \draw[thick,-stealth, cardinal, dashed] (\xp0+\barlen-2*\labelshift+\minishift,\nefflpC1 / 2 + \nefflpB2 / 2 + 0.2) -- (\xp0+\barlen+\redshift,\nefflpA2 - \redyshift - \redyoffset)  node[anchor=west]{$\Delta\beta_{02,31}$};
    
    \draw[thick, stealth-stealth, cardinal] (\xp0+\barlen-\labelshift,\nefflpA2) -- (\xp0+\barlen-\labelshift,\nefflpD1);
    \draw[thick,-stealth, cardinal, dashed] (\xp0+\barlen-1*\labelshift+\minishift,\nefflpA2 / 2 + \nefflpD1 / 2 + 0.15) -- (\xp0+\barlen+\redshift,\nefflpA2 - 0 * \redyshift - \redyoffset)  node[anchor=west]{$\Delta\beta_{21,12}$};
    
    \draw[thick, stealth-stealth, cardinal] (\xp0+\barlen-\labelshift*3,\nefflpA2) -- (\xp0+\barlen-\labelshift*3,\nefflpB2);
    \draw[thick,-stealth, cardinal, dashed] (\xp0+\barlen-3*\labelshift+\minishift,\nefflpC1 / 2 + \nefflpB2 / 2) -- (\xp0+\barlen+\redshift,\nefflpA2 - 2 * \redyshift - \redyoffset)  node[anchor=west]{$\Delta\beta_{02,12}$};
    \draw[thick, stealth-stealth, cardinal] (\xp0+\barlen-4*\labelshift,\nefflpC1) -- (\xp0+\barlen-4*\labelshift,\y0);
    \draw[thick,-stealth, cardinal, dashed] (\xp0+\barlen-4*\labelshift+\minishift,\nefflpC1 / 2 + \y0 / 2) -- (\xp0+\barlen+\redshift,\nefflpA2 - 3 * \redyshift - \redyoffset)  node[anchor=west]{$\Delta\beta_{21,\mathrm{clad}}$};
    \draw[thick, stealth-stealth, cardinal] (\xp0+\barlen-5*\labelshift,\nefflpA2) -- (\xp0+\barlen-5*\labelshift,\y0);
    \draw[thick,-stealth, cardinal, dashed] (\xp0+\barlen-5*\labelshift+\minishift,\nefflpA2 / 2 + \y0 / 2) -- (\xp0+\barlen+\redshift,\nefflpA2 - 4 * \redyshift - \redyoffset)  node[anchor=west]{$\Delta\beta_{02,\mathrm{clad}}$};
    
    \draw[solid, ultra thick, berkeley] (\xdb0,\delbeta1) -- (\xdb0+\barlen,\delbeta1) node[anchor=west]{$\Delta\beta_{01,11}$};
    \draw[solid, ultra thick, berkeley] (\xdb0,\delbeta2) -- (\xdb0+\barlen,\delbeta2) node[anchor=west]{$\Delta\beta_{11,02}$};
    \draw[solid, ultra thick, berkeley] (\xdb0,\delbeta3) -- (\xdb0+\barlen,\delbeta3) node[anchor=west]{$\Delta\beta_{11,21}$};
    \draw[solid, ultra thick, cardinal, dashed] (\xdb0,\delbeta4) -- (\xdb0+\barlen,\delbeta4) node[anchor=west, yshift=0.05cm]{$\Delta\beta_{02,31}$};
    \draw[solid, ultra thick, cardinal, dashed] (\xdb0,\delbeta5) -- (\xdb0+\barlen,\delbeta5) node[anchor=west]{$\Delta\beta_{02,12}$};
    \draw[solid, ultra thick, cardinal, dashed] (\xdb0,\delbeta6) -- (\xdb0+\barlen,\delbeta6) node[anchor=west]{$\Delta\beta_{21,31}$};
    \draw[solid, ultra thick, cardinal, dashed] (\xdb0,\delbeta7) -- (\xdb0+\barlen,\delbeta7) node[anchor=west]{$\Delta\beta_{21,11}$};
    \draw[solid, ultra thick, cardinal, dashed] (\xdb0,\delbeta8) -- (\xdb0+\barlen,\delbeta8) node[anchor=west]{$\Delta\beta_{02,\mathrm{clad}}$};
    \draw[solid, ultra thick, cardinal, dashed] (\xdb0,\delbeta9) -- (\xdb0+\barlen,\delbeta9) node[anchor=west]{$\Delta\beta_{21,\mathrm{clad}}$};
}

\tikzset{
  font={\fontsize{11pt}{12}\selectfont}}
\begin{document}
\title{Low-Loss All-Fiber Mode Permuter Design Exploiting Propagation Constant Engineering \\and Cascaded Bragg Gratings}
\author{{Oleksiy Krutko*,~\IEEEmembership{Student~Member,~IEEE}, Rebecca Refaee*,\\Anirudh Vijay,~\IEEEmembership{Student~Member,~IEEE}, Nika Zahedi, and Joseph M. Kahn,\IEEEmembership{~Fellow,~IEEE}}
\thanks{* These authors contributed equally. Oleksiy Krutko, Rebecca Refaee, Anirudh Vijay, Nika Zahedi, and Joseph M. Kahn are with the E. L. Ginzton Laboratory, Department of Electrical Engineering, Stanford University, Stanford, CA 94305 USA (email: oleksiyk@stanford.edu; becca24@stanford.edu; avijay@stanford.edu; nzahedi@stanford.edu;  jmk@ee.stanford.edu).

© 2025 IEEE. Personal use of this material is permitted. Permission from IEEE must be obtained for all other uses, in any current or future media, including reprinting/republishing this material for advertising or promotional purposes, creating new collective works, for resale or redistribution to servers or lists, or reuse of any copyrighted component of this work in other works.
}}
\markboth{Journal of Lightwave Technology}%
{Author \MakeLowercase{\textit{et al.}}: Insert Paper Title here}

\maketitle

\begin{abstract}
Managing group-delay (GD) spread is vital for reducing the complexity of digital signal processing (DSP) in long-haul systems using multi-mode fibers.  
GD compensation through mode permutation, which involves periodically exchanging power between modes with lower and higher GDs, can reduce GD spread.
GD spread is maximally reduced when the modal GDs of the transmission fiber satisfy a specific relation and the mode permuter exchanges power between specific modes.
Mode permuters with fiber Bragg gratings have been developed for links accommodating $D=6$ guided spatial and polarization modes; however, to the best of our knowledge, there have been no designs for links supporting a greater number of modes.
We present two mode permuter designs based on fiber Bragg gratings for links employing graded-index transmission fibers with $D=12$ guided spatial and polarization modes. 
One design utilizes a step-index (SI) transverse profile, and the other features a free-form-optimized transverse profile achieving improved performance.
For the SI design, we achieve mode-dependent loss standard deviation (MDL STD) and mode-averaged loss (MAL) of less than $0.14$ dB and $0.27$ dB, respectively, over the C-band.
For the design with a free-form-optimized transverse profile, we achieve MDL STD and MAL of less than $0.11$ dB and $0.07$ dB, respectively, over the C-band.
We numerically evaluate the designs through link simulations and quantify the reduction in GD spread for different levels of random inter-group coupling in the fiber. 
Our results show that in a link with periodic mode permutation and mode scrambling, the GD STD is reduced by a factor over $3.13$ compared to a link relying solely on periodic mode scrambling.
\end{abstract}

\section{Introduction}

Power-limited long-haul optical communication systems now utilize space-division multiplexing (SDM) to satisfy the demand for higher data rates. 
In current SDM systems, cable capacity is optimized under feed-power constraints by employing multiple parallel single-mode fibers (SMFs), with each fiber transmitting a lower power and data rate compared to traditional non-SDM systems  \cite{srinivas_modeling_2021, sinkin_maximum_2017, sinkin_sdm_2018, cai_9_2022}.

Coupled-core multicore fiber (CC-MCF) and multi-mode fiber (MMF) provide alternatives to SMF, offering the potential to improve integration and scalability in (SDM) systems while increasing capacity per unit fiber  \cite{klaus_advanced_2017, winzer_chapter_2013}.
Long-haul links employing MDM in MMF are attractive because they can achieve the highest level of integration \cite{winzer_chapter_2013} and can be efficiently amplified with low mode-dependent gain, while using fewer pump laser diodes per signal mode \cite{srinivas_efficient_2023}.

High-capacity long-haul MDM links have often used graded-index (GI) MMFs, owing to their relatively low uncoupled GD STD \cite{jensen_demonstration_2015,ryf_mode-multiplexed_2015}.
Within these fibers, the spatial and polarization modes form distinct mode groups. The propagation constants of modes within the same group are nearly identical, whereas those in different groups differ substantially.
Random perturbations in these fibers cause strong intra-group mode coupling and weak inter-group mode coupling \cite{fontaine_characterization_2013}.

Mode-dependent gain and loss (collectively referred to as MDL) and modal dispersion limit the performance of MDM links.
Mode scramblers are commonly utilized to introduce strong coupling among all modes to manage these effects \cite{krutko_ultra-low-loss_2025}.
Strong coupling decreases the standard deviation (STD) of link MDL, increasing average capacity and minimizing outage probability \cite{ho_mode-dependent_2011}. 
Additionally, it lowers GD STD, reducing the complexity of digital signal processing (DSP) at the receiver. 
In systems with strong coupling, GD and MDL STDs accumulate in proportion to the square root of the propagation length. 
Strong mode coupling also improves frequency diversity, which further reduces the outage probability \cite{ho_frequency_2011}.

A new method for decreasing GD STD or MDL STD by using mode permutation has recently been proposed \cite{shibahara_long-haul_2020, di_sciullo_modal_2023, wang_new_2023, wang_novel_2022, xu_modal_2023, fan_compact_2022}. 
Power is periodically exchanged between slow and fast modes or low- and high-gain modes to reduce GD STD or MDL STD accumulation, respectively. 
Experimental work has demonstrated that mode permutation can achieve GD and MDL STD accumulation in proportion to the square root of the propagation length \cite{di_sciullo_modal_2023, wang_new_2023, shibahara_long-haul_2020}, like standard strong coupling \cite{ho_mode_2013, ho_statistics_2011, ho_linear_2014, ho_mode-dependent_2011}.
The authors have shown that specific mode permutations can result in significantly less GD STD accumulation than an equivalent strongly coupled link \cite{vijay_modal_2024}. 
We refer to such a scheme as \textit{self-compensation} due to its similarity in GD accumulation to a link employing conventional GD compensation by concatenation of fibers having opposing modal GD orderings.

We will refer to the physical devices that perform mode permutation as $\it{mode}$ $\it{permuters}$.
Physical mode permuter implementations include photonic lanterns \cite{arik_group_2016}, cascaded long-period fiber Bragg gratings (LPFGs) \cite{fan_compact_2022, jin_mode_2016}, multi-plane light converters (MPLCs) \cite{di_sciullo_modal_2023, wang_new_2023}, and multiplexer/demultiplexer pairs \cite{shibahara_long-haul_2020}.
Fiber-based options like LPFGs are especially promising as they are integrable, have low device loss and high-bandwidth coupling efficiency, and are readily fabricated via ultraviolet laser exposure, $\mathrm{CO_2}$ laser irradiation, electrical discharge, femtosecond laser exposure, mechanical microbends, or etched corrugations \cite{wang_review_2010, askarov_long-period_2015, zhao_broadband_2018, ma_high-order_2023, wang_efficient_2024, wang_broadband_2024, zhao_mode_2016, zhao_mode_2017, zhao_ultra-broadband_2019}. 

As with mode scramblers, MDL and MAL requirements are strict for mode permuters since a signal must pass through tens or hundreds of these devices in a long-haul link.
Moreover, the mode permuter must operate over a wide wavelength span (e.g., the C-band) in order to obtain low GD STD and/or MDL STD for all wavelength channels.
Previously, LPFG-based mode permuters have been developed for $D=6$ spatial and polarization modes \cite{jin_mode_2016, fan_compact_2022}. 
These designs utilize a cascade of gratings that each perform power exchanges between specific pairs of modes, leveraging mode field symmetries and phase-matching.
To the best of our knowledge, an LPFG-based mode permuter for $D=12$ spatial and polarization modes has not been demonstrated and is a novel contribution of this paper. 

In this work, we study the optimal mode permutations for reducing GD STD in long-haul MDM links employing GI transmission fiber with $D=12$ spatial and polarization modes.
Our analysis shows that GD STD can be minimized by co-design of a transmission fiber and a mode permuter, in which the modal GDs of the transmission fiber must satisfy a certain relation and the mode permuter must exchange power between specific pairs of modes. 
To create a mode permuter that meets the required characteristics, we propose a design based on a cascade of LPFGs. 
Each LPFG is designed to efficiently execute a subset of the necessary mode permutations.
To simplify fabrication and minimize splicings, we limit ourselves to inscribing all gratings in the same fiber, which constrains the design of the mode permuter's transverse index profile.
We describe the requirements for the propagation constant spacings of the mode permuter fiber to enable high-efficiency mode exchanges with minimal loss and undesired mode coupling.
We present two mode permuter designs: one with a step-index (SI) transverse profile, and the other featuring a free-form-optimized transverse profile and yielding better performance. 
We examine the efficacy of our designed mode permuters in reducing link GD STD for various strengths of random inter-group coupling through numerical simulation. 
We show that combining periodic mode permutation and mode scrambling 
reduces a link's GD STD by a factor of at least $3.13$ compared to a link that relies solely on periodic mode scrambling.

The remainder of the paper is organized as follows. 
Section \ref{sec:mode_permutation_theory} describes an analytical method for computing the optimal mode permutation matrices given a fiber with specific mode groups and modal GDs.
Section \ref{sec:mode_perm_design} describes the mode permuter structure based on cascaded LPFGs, modal propagation in LPFGs, the design of the mode permuter fiber transverse index profile, and the design of the grating transverse and longitudinal index profiles. 
Section \ref{sec:results} details the two LPFG-based mode permuter designs, quantifies their performance, and evaluates the free-form-optimized mode permuter's efficacy in reducing the GD STD of a long-haul link.
Sections \ref{sec:discussion} and \ref{sec:conclusion} are the discussion and conclusion, respectively.

\section{Group Delay Management of MDM Links using Mode Permutation} \label{sec:mode_permutation_theory}

This section outlines a theoretical framework for evaluating the GD STD of self-compensated links employing a GI transmission fiber with $N_g$ mode groups and $D$ spatial and polarization modes. 
We employ this framework to first study a simple case with $N_g = 2$ mode groups and $D = 2$ modes, and then identify a self-compensation scheme that minimizes GD STD for a link with $N_g = 3$ mode groups and $D=12$ modes.

\begin{figure}
    \centering
    \includegraphics[width=0.6\linewidth]{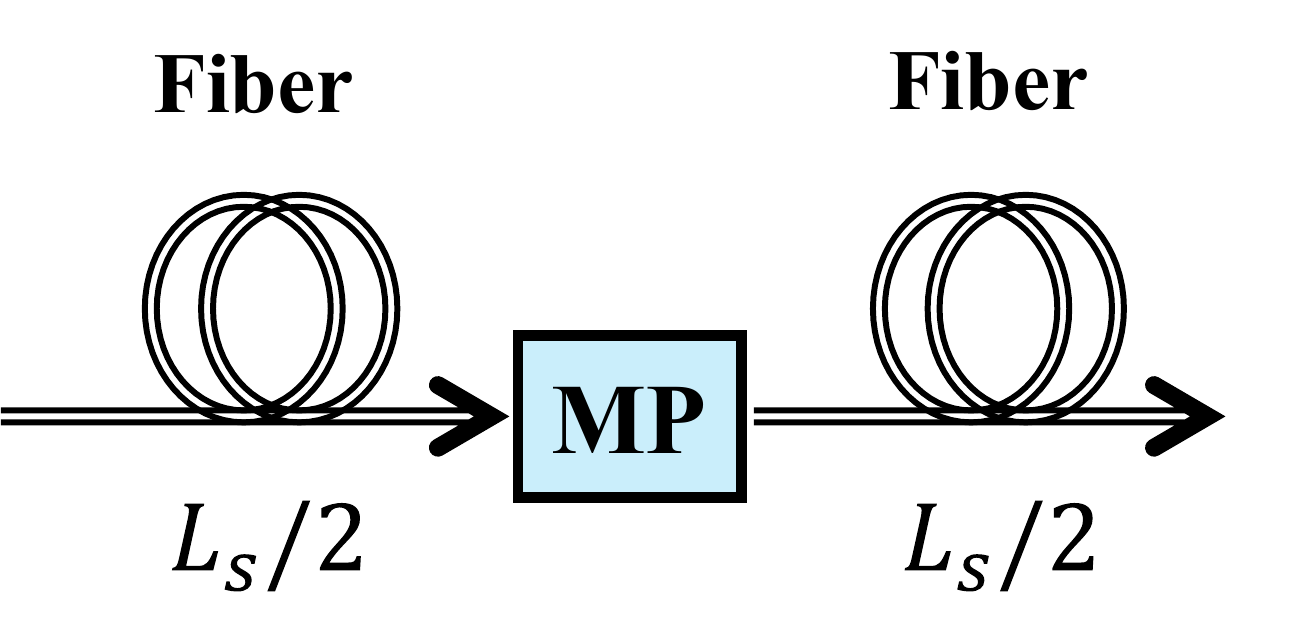}
    \caption{Diagram of self-compensation. The mode permuter is placed in between two identical fiber segments of length $L_S/2$. MP: Mode Permuter.}
    \label{fig:2SS_1MP}
\end{figure}

We consider a single span of a self-compensated link,  shown in Fig. \ref{fig:2SS_1MP}, employing a lossless, unitary mode permuter, described by a $D \times D$ transfer matrix $\RMP$, between identical fiber segments of length $L_s/2$, where $L_s$ is the length of the span. 
The $D$ modes are grouped into $N_g$ mode groups; the set of modes in the $i$th mode group is denoted by $\mathcal{M}_i$ and the degeneracy is given by $d_i=\left | \mathcal{M}_i \right|$, where $\left | \cdot \right|$ represents the cardinality. 

Following the analysis in \cite{vijay_modal_2024}, we define a mode-group-averaged $N_g$-dimensional GD vector (measured in seconds) representing each of the transmission fiber segments.  For mode group $i=1,2,\dots, N_g$, 
\begin{align*}
    \tau_{0,i} = \frac{L_{s}}{2d_i} \sum_{k\in\mathcal{M}_i}\beta_{1}[k],
\end{align*}
where $\beta_{1}[k]$ is the uncoupled GD per unit length of the $k$th mode of the transmission fiber such that $\sum_{k=1}^{D}\beta_{1}[k]=0$. 
The $N_g\times N_g$ mode-group power coupling matrix of the mode permuter is given by 
\begin{equation*}
\label{eq:PC_def}
    \PCMP[i,j] = \sum_{l\in\mathcal{M}_i}{\sum_{m \in\mathcal{M}_j}{ \left| \RMP[l,m] \right|^2}},
\end{equation*}
The $i,j$ element of $\PCMP$ is the power the mode permuter transfers from $M_i$ to $M_j$.

The GD STD of the self-compensated span is given by
\begin{align}
\begin{split}
    \sigma_{\mathrm{GD}} ={}&\sqrt{\frac{\E{\norm{\ttot}^2 }}{D}}.
\end{split}
\label{eq:gd_rms_tot}
\end{align}
In the absence of random inter-group coupling and assuming a negligible intra-group GD STD, $\E{\norm{\ttot}^2 }$ denotes the expected squared norm of the coupled GDs and is given by
\begin{equation}
    \E{\norm{\ttot}^2 } = 2\tautau_0^H\mathbf{D}\tautau_0 + 2\tautau_0^H \PCMP \tautau_0,
    \label{eq:2SS_1MP_formula}
\end{equation}
where $\mathbf{D}$ is the $N_g\times N_g$ diagonal matrix of mode group degeneracy, $\mathbf{D}[i,i] = d_i$. 
Since $\RMP$ is a unitary matrix, the constraints on the entries of $\PCMP$ are given by
\begin{equation}
    \begin{aligned}
        &\PCMP[i,j] \geq 0,~1\leq i,j \leq N_g, \\
        &\DPC_{\mathrm{MP}} \mathbf{1}_{N_g} = \mathbf{1}_{N_g},\\
        &\mathbf{d}^T \DPC_{\mathrm{MP}} = \mathbf{d}^T.\\
    \end{aligned}
    \label{eq:P_constraints}
\end{equation}
Here, $\mathbf{1}_{N_g}$ denotes the all-ones column vector of dimension $N_g$.
Finding the optimal $\PCMP$ to minimize \eqref{eq:2SS_1MP_formula} under the constraints in \eqref{eq:P_constraints} is a convex optimization problem \cite{vijay_modal_2024}. 
We obtain the optimal $\PCMP$ for two combinations of $N_g$, $D$, and $\bf{d}$.

\subsection*{Case 1: $N_g = 2$, $D = 2$, and $\mathbf{d} = [1, 1]^T$}
First, we study the case of a link supporting two modes in different mode groups.

The two components of $\tautau_0$ are defined to be equal and opposite, $\tau_{0,1} = -\tau_{0,2}$, and the constraints in \eqref{eq:P_constraints} reduce to
\begin{equation}
    \begin{aligned}
        & \PCMP[i,j] \geq 0,~1\leq i,j \leq 2, \\
        & \PC[1,1] + \PC[1,2] = \PC[1,1] + \PC[2,1] = 1, \\
        & \PC[2,1] + \PC[2,2] = \PC[1,2] + \PC[2,2] = 1, \\
    \end{aligned}
\end{equation}
We find that $\E{\norm{\ttot}^2 }=0$ when $\PCMP$ is equal to the exchange matrix $\bf{J}_2$:
\begin{equation}
    \PCMP = \bf{J}_2 = \begin{bmatrix} 0 & 1 \\ 1 & 0 \end{bmatrix}.
\end{equation} 
Intuitively, this result is expected since each signal travels equally in the slow and fast modes. 
Next, we examine a practical scenario in which varying mode-group degeneracies and uncoupled group delays complicate perfect group delay compensation.

\subsection*{Case 2: $N_g = 3$, $D = 12$, and $\mathbf{d} = [2, 4,6]^T$}
This case represents a link using a typical $D = 12$-mode GI transmission fiber with the following signal modes, each with two polarizations: $\{LP_{01}\},$ $\{ LP_{11a}$, $ LP_{11b}\},$ $\{ LP_{02},$ $ LP_{21a},$ $ LP_{21b}\}$.

Minimizing \eqref{eq:2SS_1MP_formula} requires a specific combination of $\tautau_0$ and $\PCMP$.
We find that $\E{\norm{\ttot}^2 }=0$ when
\begin{subequations}
    \begin{align}
    &\tautau_{0,1} = \tautau_{0,2} = -\tautau_{0,3}, \label{eq:opt_tau} \\ 
    &\PCMP = 2\begin{bmatrix}
        0 & 0 & 1 \\ 
        0 & 0 & 2 \\
        1 & 2 & 0
    \end{bmatrix}. \label{eq:opt_tau_and_P}
    \end{align}
    
\end{subequations}
This combination of $\tautau_0$ and $\PCMP$ requires the GDs of the first two mode groups to be equal and opposite to those of the third mode group, while the mode permuter must exchange all power between the third mode group and the first two mode groups.
In the next section, we propose and design an LPFG-based mode permuter with the optimal mode-averaged power coupling matrix in \eqref{eq:opt_tau_and_P}.

The self-compensation schemes outlined here ideally achieve total GD compensation, but their implementation is challenging due to unwanted inter-group coupling arising from link components, splicing, and fiber manufacturing defects. 
This unwanted mode coupling hinders equal transmission of signals in both slow and fast modes, thereby compromising the GD STD reduction. 
These same issues also impact conventional GD compensation \cite{arik_delay_2015}.
In Section \ref{sec:results}, when assessing the performance of our designed mode permuters in a self-compensated link, we account for the influence of splicing and random inter-group coupling.

\section{Mode Permuter Design Methodology}
\label{sec:mode_perm_design}
This section presents a methodology for the design of a $D=12$-mode LPFG-based mode permuter that exchanges all power between the third mode group and the first two mode groups.
In the next section, we employ this methodology to design LPFG-based mode permuters for a long-haul MDM system supporting $D = 12$ guided spatial and polarization modes, which we refer to as the signal modes.

We describe the mode permuter design methodology primarily with scalar, linearly polarized modes, making reference to vector modes when necessary.
The term ``mode permuter" hereafter refers to LPFG-based mode permuters unless noted otherwise.

\begin{figure}
    \centering
    \includegraphics[width=1\linewidth]{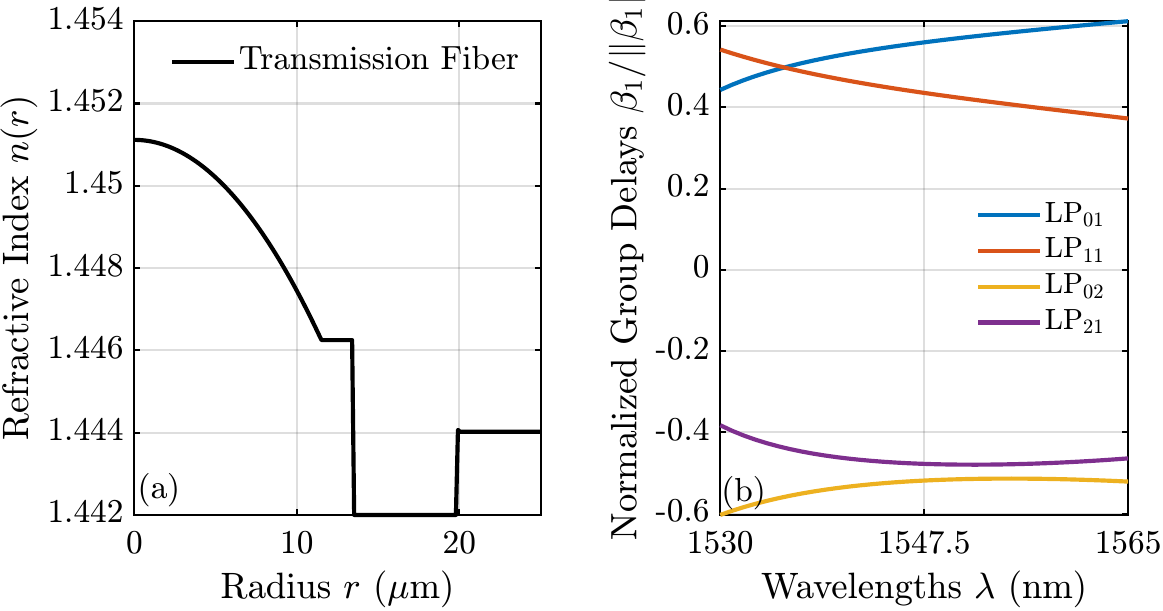}
    \caption{(a) Transverse index profile and (b) normalized group delays of the six-spatial-mode transmission fiber designed for self-compensation.}
    \label{fig:transmission_fiber}
\end{figure}

We use the methodology described in \cite{vijay_closed-form_2025} to design the link transmission fiber.
Fig. \ref{fig:transmission_fiber} shows the transverse index profile and normalized group delays of the transmission fiber.
The transverse profile is a modified GI fiber incorporating both a pedestal and a trench to obtain mode-group averaged GDs approximately satisfying \eqref{eq:opt_tau} and sufficiently low bending losses. 
While the modal dispersions are not ideal, we believe this fiber represents a realistic scenario.

\subsection{Proposed Mode Permuter Structure and Design}
\label{subsec:mode_permuter_structure}

\label{subsec:proposed-struct}
\begin{figure*}[htbp]
    \centering
    \includegraphics[width=0.9\linewidth]{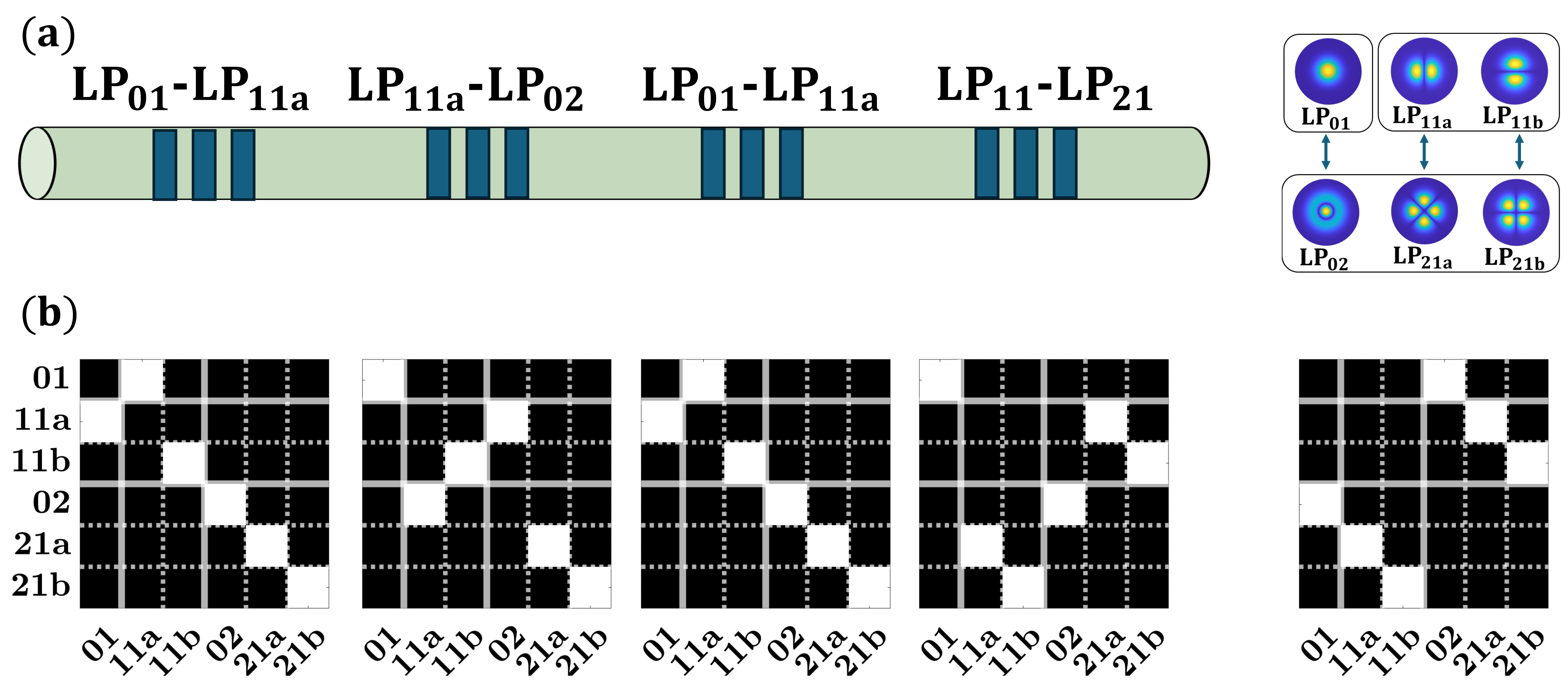}
    \caption{(a) Principal structure of the mode permuter, comprised of four cascaded LPFGs, and the resulting mode exchanges. (b) The transfer matrices of the individual gratings and the mode permuter, shown in the basis of LP spatial modes. The first three gratings exchange $\LP_{01}$ and $\LP_{02}$ with $\LP_{11a}$ as an intermediate stage. The final grating exchanges $\LP_{11}$ and $\LP_{21}$. }
    \label{fig:grating_diagram}
\end{figure*}

Numerous mode permutations can yield the optimal mode-group-summed power coupling matrix $\PCMP$ in \eqref{eq:opt_tau_and_P}.
Due to differing mode field symmetries, exchanging power between $\LP_{01}$ and $\LP_{02}$ and between $\LP_{11}$ and $\LP_{21}$ is best.
In the spatial mode basis $\{\LP_{01}, \LP_{11a}, \LP_{11b}, \LP_{02}, \LP_{21a}, \LP_{21b} \}$, the transfer matrix of these mode permutations is given by
\begin{equation}
    {\RMP} = \begin{bmatrix}  
        0 & 0 & 0 & \bf{1} & 0 & 0 \\ 
        0 & 0 & 0 & 0 & \bf{1} & 0 \\
        0 & 0 & 0 & 0 & 0 & \bf{1} \\
        \bf{1} & 0 & 0 & 0 & 0 & 0 \\
        0 & \bf{1} & 0 & 0 & 0 & 0 \\
        0 & 0 & \bf{1} & 0 & 0 & 0 
        \end{bmatrix}.
\end{equation}
Since only the power is relevant, the phase of each entry can be arbitrary as long as $\RMP$ remains unitary.

There are two approaches to obtaining this transfer matrix using cascaded gratings. 
One possibility is to cascade two gratings: $\{\LP_{01}$\textendash
$\LP_{02}$, $\LP_{11}$\textendash$\LP_{21}\}$.
Another possibility is to cascade four gratings: $\{\LP_{01}$\textendash$\LP_{11a}$, $\LP_{11a}$\textendash$\LP_{02}$, $\LP_{01}$\textendash$\LP_{11a}$, $\LP_{11}$\textendash$\LP_{21}\}$. In this case, the first three gratings perform the $\LP_{01}$ and $\LP_{02}$ power exchange by transferring power through $\LP_{11a}$. 

We considered both options and found that the four-grating cascade performs best.
Although the two-grating cascade uses fewer gratings, the $\LP_{01}$\textendash$\LP_{02}$ grating requires a mode permuter fiber with a core-cladding index difference significantly larger than that of the transmission fiber to prevent the grating from also coupling the $\LP_{21}$ and $\LP_{02}$ modes into unguided modes. 
The four-grating cascade solely transfers power between modes of adjacent mode groups, so the core-cladding index difference can be lower.
As a result, the splicing loss between the mode permuter fiber and the transmission fiber is significantly lower for the four-grating cascade than for the two-grating cascade.
One disadvantage of the four-grating cascade is that the efficiencies of the $\LP_{01}$\textendash$\LP_{11a}$ and $\LP_{11a}$\textendash$\LP_{02}$ mode conversions are degraded by the instability of the $\LP_{11a,x}$ and $\LP_{11a,y}$ modes, which periodically convert to $\LP_{11b,y}$ and $\LP_{11b,x}$, respectively, according to the beat lengths of the $\mathrm{TE}_{01}$/$\mathrm{HE}_{21}$ and $\mathrm{TM}_{01}$/$\mathrm{HE}_{21}$ vector mode pairs that compose the $\LP_{11}$ modes \cite{kogelnik_modal_2012, wang_broadband_2024, wang_efficient_2024}. 
Although this impairs the mode conversion efficiency, we find that the reduced splicing loss is a preferable trade-off.

Fig. \ref{fig:grating_diagram} illustrates the key components of the proposed mode permuter, consisting of four cascaded LPFGs. 
The first three gratings facilitate the exchange between $\LP_{01}$ and $\LP_{02}$, using $\LP_{11a}$ as an intermediate state. 
The last grating exchanges both $\LP_{11}$ modes with both $\LP_{21}$ modes. 
The ideal transfer matrices for the individual gratings, as well as their combination, are shown in the basis of LP spatial modes.

To implement the proposed mode permuter, we must design three distinct gratings: $\LP_{01}$\textendash$\LP_{11a}$, $\LP_{11a}$\textendash$\LP_{02}$, and $\LP_{11}$\textendash$\LP_{21}$. 
We limit our designs to those using the same fiber for all gratings to reduce the number of splices needed. 
This choice makes designing the mode permuter fiber's transverse index profile more challenging, yet simplifies its fabrication.

\subsection{Modeling Methodology}
\label{subsec:modeling_methodology}

The total refractive index (RI) for each of the gratings in the mode permuter can be written as 
\begin{equation}
    n_\mathrm{tot}(r,\theta,z) = n(r) + \Delta n_{\mathrm{g}}(r,\theta) \left( 1 + \cos \left( \frac{2\pi z}{\Lambda}+ \phi(z)  \right) \right)
    \label{eq:grating_index}
\end{equation}
where $n(r)$ is the mode permuter fiber's transverse index profile, $\Delta n_{\mathrm{g}}(r,\theta)$ is the grating transverse index profile change induced by single- or multi-sided UV illumination, and $\phi(z)$ is the longitudinal chirp profile.
$n(r)$ is not necessarily the same as the transverse index profile of the transmission fiber $n_\mathrm{init}(r)$. 
We assume the index modulation induced by single-sided UV exposure is present only in the fiber core region and follows a decreasing exponential function, where it is given by
\begin{equation}
\begin{split}
    \Delta n_{\mathrm{ss}}(r,\theta;\psi) = &\chi \exp \biggl[-\rho \biggl( \sqrt{r_{\mathrm{core}}^2 - r^2\sin^2{(\theta - \psi)}} \\ &\qquad - r\cos{(\theta - \psi)}   \biggr)   \biggr] P(r)
\end{split}
\label{eq:grating-transverse}
\end{equation}
where $\chi$ is the modulation depth, $\rho$ is the exponential decay rate of the index modulation, $r_{\mathrm{core}}$ is the core radius, $\psi$ is the azimuthal angle of the UV illumination, and $P(r)$ is the radial dependence of the index modulation \cite{fang_low-dmd_2015, lu_full_2010}. Here, $P(r)=1$ within the core and $P(r)=0$ outside the core. 
The grating transverse index profile is directional and axially asymmetric, enabling coupling of modes with differing angular symmetries \cite{fang_low-dmd_2015}.

The coupled-mode propagation equations describe the propagation of both guided and unguided modes in the mode permuter and can be solved to obtain $\RMP$ \cite{askarov_long-period_2015, lu_full_2010, fang_low-dmd_2015}.
The guided and unguided mode fields and propagation constants $\bvec$ are computed in a radially resolved cylindrical geometry with a perfectly matched layer and zero termination boundary condition, as in \cite{askarov_long-period_2015, lu_full_2010}. 
We assume the outer cladding is index-matched to the inner cladding, which can be achieved with an index-matched coating \cite{hale_optical_2001}; thus, no discrete cladding modes are supported \cite{askarov_long-period_2015} and inner-outer cladding reflections are suppressed \cite{hale_optical_2001}. 
Loss in $\RMP$ arises from undesired coupling between guided signal modes and unguided modes or guided non-signal modes during the permutation process. 
We also consider loss and coupling from modal field mismatches between the modes in the transmission fiber and those in the mode permuter, which are referred to as splicing losses \cite{meunier_efficient_1991}.
Combining all these effects, we can evaluate the mode permuter MDL STD and MAL using the modal gain operator \cite{ho_mode-dependent_2011}.

\subsection{Mode Permuter Fiber Transverse Index Design}
\label{subsec:prop_const_eng}

In this subsection, we first explain key constraints for the modal propagation constants of the mode permuter fiber. 
We then explain how to design an SI fiber that satisfies these constraints and how to use free-form index optimization to obtain an improved design.

\subsubsection{Working Principle of LPFGs}
Mode coupling by an LPFG is a coherent, phase-matched process \cite{askarov_long-period_2015, fang_low-dmd_2015, liu_reducing_2018}. 
A uniform grating couples two modes most efficiently when
\begin{equation}
    \Delta\beta=\frac{2\pi}{\Lambda},
    \label{eqn:phase-matching}
\end{equation}
where the propagation constant spacing $\Delta\beta$ is the difference between the modal propagation constants and $\Lambda$ is the grating period.
Considering a simple case in which an MMF supports only two modes, we can evaluate the maximum coupling efficiency $\eta$ between the modes achievable by a grating: 
\begin{equation}
    \eta \approx \frac{1}{1 + \left( \frac{\Delta\beta-2\pi/\Lambda}{\kappa} \right) ^2},
    \label{eqn:coupling_efficiency}
\end{equation}
where $\kappa$ is the coupling coefficient between the two modes induced by the grating and the self-coupling coefficients are assumed to be negligible \cite{fang_low-dmd_2015}.
The grating length for greatest power transfer occurs at length
\begin{equation*}
    L_c = \frac{\pi}{2|\kappa|}.
\end{equation*}
For coupling between mode $i$, $\LP_{lm}$, and mode $j$, $\LP_{l'm'}$, we denote the propagation constant spacing as:
\begin{equation*}
    \Delta\beta_{lm,l'm'} = |\beta[i] - \beta[j]|.
\end{equation*}

\subsubsection{Design Criterion}
Mode permutation relies on controlled, complete power exchanges between particular pairs of modes. 
To achieve this, we require the $D=12$ mode permuter fiber to satisfy the following design criterion:
\begin{enumerate}
    \item [C.1]
    The propagation constant spacings of the desired mode exchanges $\LP_{01}$\textendash$\LP_{11}$, $\LP_{11}$\textendash$\LP_{02}$, and $\LP_{11}$\textendash$\LP_{21}$ must be well separated from each other and from every spacing between a signal mode and any other signal, guided non-signal, or unguided mode.  
\end{enumerate}
If this design criterion is satisfied, the $\LP_{01}$\textendash$\LP_{11a}$, $\LP_{11a}$\textendash$\LP_{02}$, and $\LP_{11}$\textendash$\LP_{21}$ gratings can be inscribed in the same mode permuter fiber, each performing their desired mode exchanges with high coupling efficiency while exhibiting low loss and minimal unwanted coupling.

\subsubsection{Choice of Mode Permuter Fiber}
\label{subsec:choice_of_fiber}
A GI-MMF, which has been suggested for mode scramblers 
\cite{askarov_long-period_2015, liu_reducing_2018, krutko_ultra-low-loss_2025} or long-haul transmission fibers \cite{srinivas_efficient_2023, di_sciullo_modal_2023, sillard_low-differential-mode-group-delay_2016}, is inadequate for our proposed mode permuter. 
This inadequacy arises because the propagation constant spacings between modes of adjacent mode groups are nearly identical, and the $\LP_{02}$ and $\LP_{21}$ modes are nearly degenerate \cite{krutko_ultra-low-loss_2025}. 
As a result, it becomes impossible to couple specific pairs of modes without unintentionally coupling others. 
In contrast, SI-MMFs do not have these properties and thus are a more suitable option.

\begin{figure*}
    \centering
    \scalebox{0.9}
        {
            \begin{tikzpicture}
                \drawStepIndexAnalysis
            \end{tikzpicture}
            
        }
    \caption{Example of an SI fiber suitable for the mode permuter described in Section \ref{subsec:mode_permuter_structure}: (a) transverse index profile, (b) propagation constants and (c) propagation constant spacings between modes in adjacent mode groups. In (b), guided modes not used for signal propagation and the cladding index are in gray. In (b) and (c), propagation constant spacings for the $\LP_{01}\textrm{\textendash}\LP_{11}$, $\LP_{11}\textrm{\textendash}\LP_{02}$, and $\LP_{11}$\textendash$\LP_{21}$ transitions are in blue, while all others are dashed and gray.}
    
    \begin{subfigure}{0\textwidth}
        \phantomcaption{\label{subfig:ref_ind}}
    \end{subfigure}
    \begin{subfigure}{0\textwidth}
        \phantomcaption{\label{subfig:prop_levels}}
    \end{subfigure}
    \begin{subfigure}{0\textwidth}
        \phantomcaption{\label{subfig:spacing_levels}}
    \end{subfigure}
    \label{fig:Step_Index_Diagram}
\end{figure*}
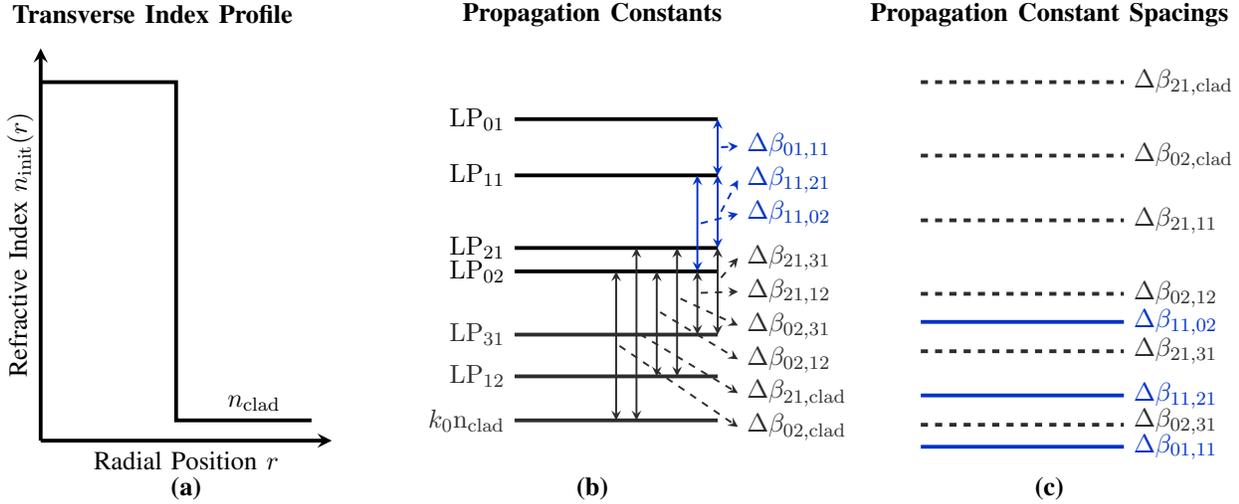

Fig. \ref{fig:Step_Index_Diagram} illustrates an exemplary SI transverse index profile with propagation constants and propagation constant spacings that satisfy the design criterion C.1.
Fig. \ref{subfig:ref_ind} shows the SI transverse index profile. Fig. \ref{subfig:prop_levels} and Fig. \ref{subfig:spacing_levels} show the propagation constants and the corresponding propagation constant spacings, respectively.
In Fig. \ref{subfig:spacing_levels}, the spacings of desired mode transitions, identified in C.1, are in blue while all others are in gray.
Visually, design criterion C.1 requires that the propagation constant spacings of the desired transitions (blue) are not near each other or other transitions (gray).

We consider two transverse index profile designs. 
In the first design, we restrict ourselves to SI profiles. 
In the second design, we use free-form index optimization from \cite{choutagunta_designing_2021} to find an improved design.
For both fibers, we follow two objectives to guide our design process. Firstly, we want to minimize the MAL and MDL STD from splicing between the transmission fiber and the mode permuter fiber to reduce device loss. Secondly,  we want to maximize the separation between the desired transitions and adjacent transitions to minimize unwanted coupling to guided or unguided modes.
\subsubsection{Step-index Profile Optimization}

For the SI design, we only need to find a combination of core radius $r_{\mathrm{core}}$ and relative index difference $\Delta$. 
We search for designs that have a propagation constant spacing difference of over 800 $\mathrm{m}^{-1}$ between desired transitions and their adjacent transitions. 
In the search, we exclude the inefficient $\LP_{02}\textrm{\textendash}\LP_{31}$ transition, as its coupling coefficient is an order of magnitude smaller than that of other transitions for the UV illuminations we use\footnote{The coupling coefficient is weak primarily because of the large mismatch in mode angular symmetry.}.
 A separation of only 100 $\mathrm{m}^{-1}$ is sufficient to effectively suppress this mode exchange.
From these designs, we select a design with the lowest MDL STD and MAL.

\subsubsection{Free-form Index Optimization}

We perform free-form optimization to design a fiber that is an improvement from the SI fiber in two regards, the first being lower MDL and MAL when splicing to and from the transmission fiber. To that end, the transmission fiber is used as the initial design that is perturbed to achieve the final mode permuter fiber. 

The second objective is to further separate the propagation constant spacings of the desired transitions from all other spacings to limit any unwanted coupling. Since the SI fiber is much closer to meeting this objective than the transmission fiber, the SI fiber is the target in the first optimization phase. 

We use the propagation constant engineering method described in \cite{choutagunta_designing_2021}. The objective function is a weighted sum of the squared differences between the actual propagation constants $\boldsymbol{\beta}$ and target propagation constants $\boldsymbol{\beta}^{\mathrm{tar}}$:
\begin{equation}
    J = \sum_{i\in\mathcal{M}_\mathrm{guided}} w_i\left( \beta_{i}-\beta_{i}^{\mathrm{tar}} \right)^2 ,
    \label{eq:ri_bojective}
\end{equation}
where $\mathcal{M}_\mathrm{guided}$ is the set of guided mode indices and $w_i$ is the weighting placed on mode $i$. The weighting parameter is determined heuristically for each phase of the optimization, with greater emphasis placed on modes involved in transitions that are more challenging to space apart from others. We use gradient descent to iteratively update the transverse index profile at each radial coordinate $r$ as 

\begin{equation}
    n(r) \leftarrow n(r)-\mu\frac{\partial J}{\partial n(r)},
\end{equation}
where $\mu$ is a step-size parameter that is chosen sufficiently small that perturbative modeling is valid.  
Details on computing the derivative $\frac{\partial J}{\partial n(r)}$ can be found in \cite{choutagunta_designing_2021}.

In subsequent optimization phases, the design is further improved by making the target propagation constants more idealized than those of the SI fiber, allowing the new design to surpass its performance. The target spacings are spread apart further and, in some phases, strategically rearranged to create more separation from the desired transitions. Multiple optimization phases, each targeting specific transitions, help ensure the design is maximally optimized. These additional phases continue to fine-tune the target spacings until the desired transitions are sufficiently isolated.

The optimization process features a tradeoff between maintaining low splicing loss and achieving a large separation between desired and undesired transitions. Perturbations to the RI profile must remain subdued to limit the loss from splicing between the mode permuter and the transmission fiber. Meanwhile, the transmission fiber's equally spaced propagation constants must be changed rather dramatically to overcome the inadequacies enumerated in Section \ref{subsec:choice_of_fiber}.

\subsection{Grating Transverse and Longitudinal Index Design}
\label{subsec:grating_design}

{\renewcommand{\arraystretch}{1}
\setlength{\tabcolsep}{0.2em}
\newcolumntype{?}{!{\vrule width 1pt}}
\begin{table*}
\centering
\begin{threeparttable}
\caption{Coupling Coefficients for Single-sided and Double-sided UV Illumination}
\begin{tabular}{c?cc|cc|cccc}
\toprule
Transitions & $\LP_{01}$\textendash$\LP_{11a}$  &  $\LP_{01}$\textendash$\LP_{11b}$  & $\LP_{11a}$\textendash$\LP_{02}$  &
$\LP_{11b}$\textendash$\LP_{02}$  &
$\LP_{11a}$\textendash$\LP_{21a}$ &
$\LP_{11b}$\textendash$\LP_{21b}$ &
$\LP_{11a}$\textendash$\LP_{21b}$ &
$\LP_{11b}$\textendash$\LP_{21a}$ \\ \hline
\begin{tabular}{@{}c@{}} Single-sided \\ $\chi = 15 \times 10^{-5}$, $\psi = 0$ \end{tabular} &
$43.1$ & $0$ & $32.0$ & $0$ & ${35.6}$ & ${40.5}$ & $0$ & $0$ \\ \hline \begin{tabular}{@{}c@{}} Double-sided \\ $\chi = 15 \times 10^{-5}$, $\psi=\pm\frac{\pi}{6}$\end{tabular} &
$37.3$ & $0$ & $27.7$ & $0$ & ${32.9}$ & ${32.9}$ & $0$ & $0$  \\
\bottomrule
\end{tabular}
The coupling coefficients between select signal modes induced by single- or double-sided UV illumination. Some coupling coefficients are exactly $0$ due to the symmetries of the modes and the grating transverse index profile..
\label{table:coupling_coeffs}
\end{threeparttable}
\end{table*}
}

This subsection describes the design of the transverse and longitudinal index profiles of the $\LP_{01}$\textendash$\LP_{11a}$, $\LP_{11a}$\textendash$\LP_{02}$, and $\LP_{11}$\textendash$\LP_{21}$ gratings.

Given a transverse index profile with propagation constant spacings satisfying design criterion C.1, the design of each grating follows the same phase-matching principles relevant to the design of low-loss and large bandwidth LPFG-based mode converters described in references \cite{wang_efficient_2024, wang_broadband_2024, zhao_mode_2017, ma_high-order_2023}.
We restrict ourselves to chirped, sinusoidal gratings described by \eqref{eq:grating_index}. 
We assume that the transverse index perturbation induced by a single-sided UV illumination is given by \eqref{eq:grating-transverse}. 
Each grating will be inscribed with one or more UV illuminations.
Therefore, for each grating we need to determine the transverse index profile and grating parameters: modulation depth $\chi$, UV illumination angle(s) $\psi$, grating period $\Lambda$, grating length $L_g$, and chirp profile $\phi(z)$. 

\subsubsection{Grating Transverse Index Profile}
First, we determine the transverse index profile for each of the gratings.
The $\LP_{01}$\textendash$\LP_{11a}$ and $\LP_{11a}$\textendash$\LP_{02}$ gratings involve coupling an axially symmetric LP mode $\LP_{0m}$ to the axially asymmetric $\LP_{11a}$. 
Since the transverse perturbation induced by a single-sided UV illumination is directional and axially asymmetric, we can independently couple $\LP_{01}$ or $\LP_{02}$ to the $a$ or $b$ modes of $\LP_{11}$ by setting the UV illumination angle $\psi$ to $0$ or $\frac{\pi}{2}$, respectively.
A single-sided grating is not suitable for exchanging all power between $\LP_{11a}$\textendash$\LP_{21a}$ and $\LP_{11b}$\textendash$\LP_{21b}$ because its transverse index profile results in unequal coupling coefficients for those pairs of modes\cite{wang_broadband_2024, wang_efficient_2024}.
We can resolve this issue by using double-sided UV illumination \cite{wang_broadband_2024}.
We find that a double-sided UV illumination at angles $\psi=\frac{\pi}{6}$ and $\psi=-\frac{\pi}{6}$ and identical modulation depths will equalize those coupling coefficients.
In this case, the transverse index perturbation is given by
\begin{equation}
    \Delta n_{\mathrm{ds}}(r,\theta) = 
    \Delta n_{\mathrm{ss}}(r,\theta;\psi=\frac{\pi}{6}) + \Delta n_{\mathrm{ss}}(r,\theta;\psi=-\frac{\pi}{6})
\label{eq:double-illum-grating-transverse}
\end{equation}

Table \ref{table:coupling_coeffs} shows the coupling coefficients $\kappa$ between the $\LP_{01}$ and $\LP_{11}$ modes, $\LP_{02}$ and $\LP_{11}$ modes, and $\LP_{11}$ and $\LP_{21}$ modes for a single-sided UV illumination grating described by $\Delta n_{\mathrm{ss}}(r,\theta;\psi=0)$ and a double-sided UV illumination described by $\Delta n_{\mathrm{ds}}(r,\theta)$.
The single-sided illumination uses a modulation depth of $\chi = 15 \times 10^{-5}$, while the double-sided illumination uses a modulation depth of $\chi = 7.5 \times 10^{-5}$.
Some coupling coefficients are exactly $0$ due to the symmetries of the modes and the UV illumination transverse profile.

Table \ref{table:coupling_coeffs} verifies that a single-sided illumination can couple $\LP_{01}$ or $\LP_{02}$ to $\LP_{11a}$ without simultaneously coupling to $\LP_{11b}$ and is therefore suitable for the $\LP_{01}$\textendash$\LP_{11a}$ and $\LP_{11a}$\textendash$\LP_{02}$ gratings.
Performing these couplings with a double-sided illumination offers no advantage.
We also confirm that a single-sided illumination induces unequal coupling coefficients between $\LP_{11a}$\textendash$\LP_{21a}$ and $\LP_{11b}$\textendash$\LP_{21b}$, while the double-sided UV illumination in \eqref{eq:double-illum-grating-transverse} equalizes those coupling coefficients. 
Hence, we will use single-sided illumination for the $\LP_{01}$\textendash$\LP_{11a}$ and $\LP_{11a}$\textendash$\LP_{02}$ gratings and double-sided illumination for the $\LP_{11}$\textendash$\LP_{21}$ grating.

\subsubsection{Grating Longitudinal Index Profile}
We now outline the design of the grating longitudinal profile. 
A linearly chirped grating is suitable for achieving high coupling efficiency across the entire C-band, as it offers greater bandwidth than a uniform grating due to enhanced phase-matching. 
The design of linearly chirped mode converters has been described in \cite{ostling_broadband_1996}.

Following \cite{ostling_broadband_1996}, we use a linear chirp profile given by 
\begin{equation}
    \phi(z) = \frac{\alpha z^2}{2L_c^2 }.
\end{equation}
where $\alpha$ is the chirp parameter.
Therefore, the grating period changes along the grating's length and is expressed by
\begin{equation}
    \frac{1}{\Lambda(z)} = \frac{1}{\Lambda_0} - \frac{\alpha z}{2\pi L_c ^2},
\end{equation}
where $\Lambda_0$ is the grating period at $z=0$.
Depending on the chirp parameter $\alpha$, there is an optimal grating length $L_g$ for high bandwidth mode conversion, which has been detailed in \cite{ostling_broadband_1996}.
For maximum coupling efficiency at the center wavelength $\lambda_0$, $\Lambda_0$ can be evaluated by
\begin{equation}
    \frac{1}{\Lambda_0} = \frac{\Delta\beta_{lm,l'm'}(\lambda_0)}{2\pi} + \frac{\alpha L_g}{4\pi L_c ^2}.
    \label{eq:get_lambda0}
\end{equation}
Finally, the modulation depth $\chi$ can be adjusted to meet coupling efficiency targets.
\section{Results}
\label{sec:results}
In this section, we apply our proposed design strategy to design two LPFG-based mode permuters for a long-haul MDM link that supports $D=12$ spatial and polarization modes. One design uses an SI transverse index profile for the mode permuter fiber, while the other uses a free-form-optimized profile. Finally, we evaluate the GD STD of a free-form-optimized mode permuter in a long-haul MDM link employing self-compensation. All values that depend on wavelength are provided over the C-band unless stated otherwise. 

\subsection{Design Case 1: Step-index Fiber}
\label{subsec:design_case_step}

 \begin{figure*}[!ht]
    \centering
    \includegraphics[width=1\linewidth]{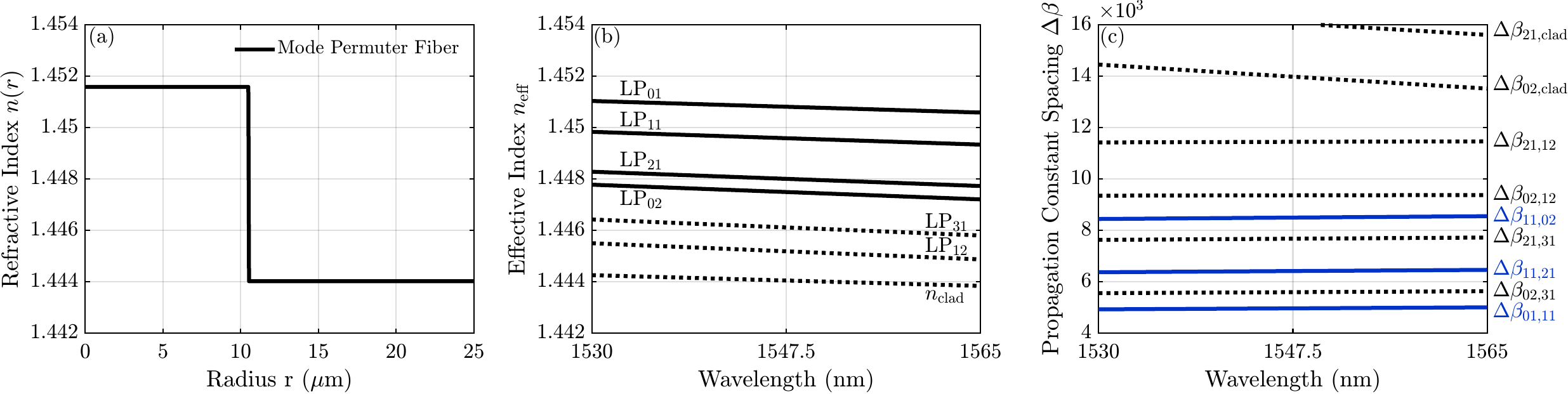}
    \caption{SI-fiber mode permuter: (a) Transverse index profile, (b) guided mode effective indices and cladding index as a function of wavelength, and (c) select propagation constant spacings as a function of wavelength. In (b), guided modes not used for signal propagation and the cladding index are dashed. In (c), propagation constant spacings for the $\LP_{01}$\textendash$\LP_{11a}$, $\LP_{11a}$\textendash$\LP_{02}$, and $\LP_{11}$\textendash$\LP_{21}$ transitions are in blue, while all others are dashed.}
    \label{fig:step_index_design}
\end{figure*}

Following the methodology in Section \ref{sec:mode_perm_design}, we design a mode permuter with gratings inscribed on an SI-MMF. 
We find the optimal core radius and relative index difference using a grid search.
The core radius $r_{\mathrm{core}}$ is varied between $7$ $\mu$m and $13$ $\mu$m, while the relative index difference $\Delta$ is varied between $0.4\%$ and $0.7\%$.
Each design is evaluated for splicing loss to the transmission fiber and maximum separation of the $\LP_{01}\textrm{\textendash}\LP_{11}$, $\LP_{11}\textrm{\textendash}\LP_{02}$, and $\LP_{11}\textrm{\textendash}\LP_{21}$ transitions from other transitions.

We obtain optimal SI-MMF parameters $r_{\mathrm{core}}=10.5$ $\mu$m and $\Delta=0.52\%$.
Fig. \ref{fig:step_index_design} illustrates the optimized SI fiber's transverse index profile, effective indices, and propagation constant spacings.
The splicing MAL and MDL STD are under $0.08$ dB and $0.05$ dB, respectively. 
The propagation constant spacings $\Delta\beta_{01,11}$, $\Delta\beta_{11,02}$, and $\Delta\beta_{11,21}$ vary over wavelength by $77.5$ $\mathrm{m}^{-1}$, $109.3$ $\mathrm{m}^{-1}$, and $91.8$ $\mathrm{m}^{-1}$, respectively.
The separations between the $\LP_{01}$\textendash$\LP_{11}$, $\LP_{11}$\textendash$\LP_{02}$, and $\LP_{11}$\textendash$\LP_{21}$ transitions and adjacent transitions are greater than $800$ $\mathrm{m}^{-1}$, excluding the inefficient $\LP_{02}\textrm{\textendash}\LP_{31}$ transition.
The beat lengths of the $\TE_{01}/\HE_{21}$ and $\TM_{01}/\HE_{21}$ vector modes,  which compose the $\LP_{11a,x}$ and $\LP_{11a,y}$ modes, are about $0.53$ $\mathrm{m}$ and $1.23$ $\mathrm{m}$, respectively.
\setlength{\tabcolsep}{2pt}
{\renewcommand{\arraystretch}{1.2}
\begin{table}
\centering

\begin{threeparttable}
\caption{Grating Parameters for Step-index Fiber}
\begin{tabular}{c|c|c|c}
\toprule
Grating & $\LP_{01}$\textendash$\LP_{11a}$  & $\LP_{11a}$\textendash$\LP_{02}$ & $\LP_{11}$\textendash$\LP_{21}$ \\ \hline 
{UV Illumination Type} & Single & Single & Double \\
{UV Illumination Angle(s) $\psi$} & $0$ & $0$ & ${\pi}/{6}$ and $-{\pi}/{6}$ \\
{Modulation Depth $\chi$} & $40\times 10^{-5}$ & $30\times 10^{-5}$ & {$10\times 10^{-5}$}\\
{Grating Period $\Lambda$ ($\mu$m)} & $1181.1$ & $717.6$ & {$948.5$}\\
{Grating Length $L_g$ (cm)} & $2.75$ & $5.01$ & {$7.29$}\\
{Chirp Parameter $\alpha$} & 4 & 4 & 4 \\
\bottomrule
\end{tabular}

\label{table:step_index_grating_params}
\end{threeparttable}
\end{table}
}

The $\LP_{01}$\textendash$\LP_{11a}$, $\LP_{11a}$\textendash$\LP_{02}$, and $\LP_{11}$\textendash$\LP_{21}$ gratings are designed according to to the procedure in \ref{subsec:grating_design}.
Table \ref{table:step_index_grating_params} provides the UV illumination type, UV illumination angle(s), and values of the modulation depth $\chi$, grating period $\Lambda$, grating length $L_g$, and chirp parameter $\alpha$ for the $\LP_{01}$\textendash$\LP_{11a}$, $\LP_{11a}$\textendash$\LP_{02}$, and $\LP_{11}$\textendash$\LP_{21}$ gratings designed using the SI fiber.
All gratings have a chirp parameter $\alpha=4$ to increase coupling efficiency over the C-band. 
The initial grating period $\Lambda_0$ is set according to \eqref{eq:get_lambda0} with a center wavelength $\lambda_0 = 1547.5$ nm.
The modulation depth of each grating is varied to obtain C-band coupling efficiency of over $97\%$ while keeping MAL below $0.1$ dB.

\begin{figure*}
    \centering
    \includegraphics[width=1\linewidth]{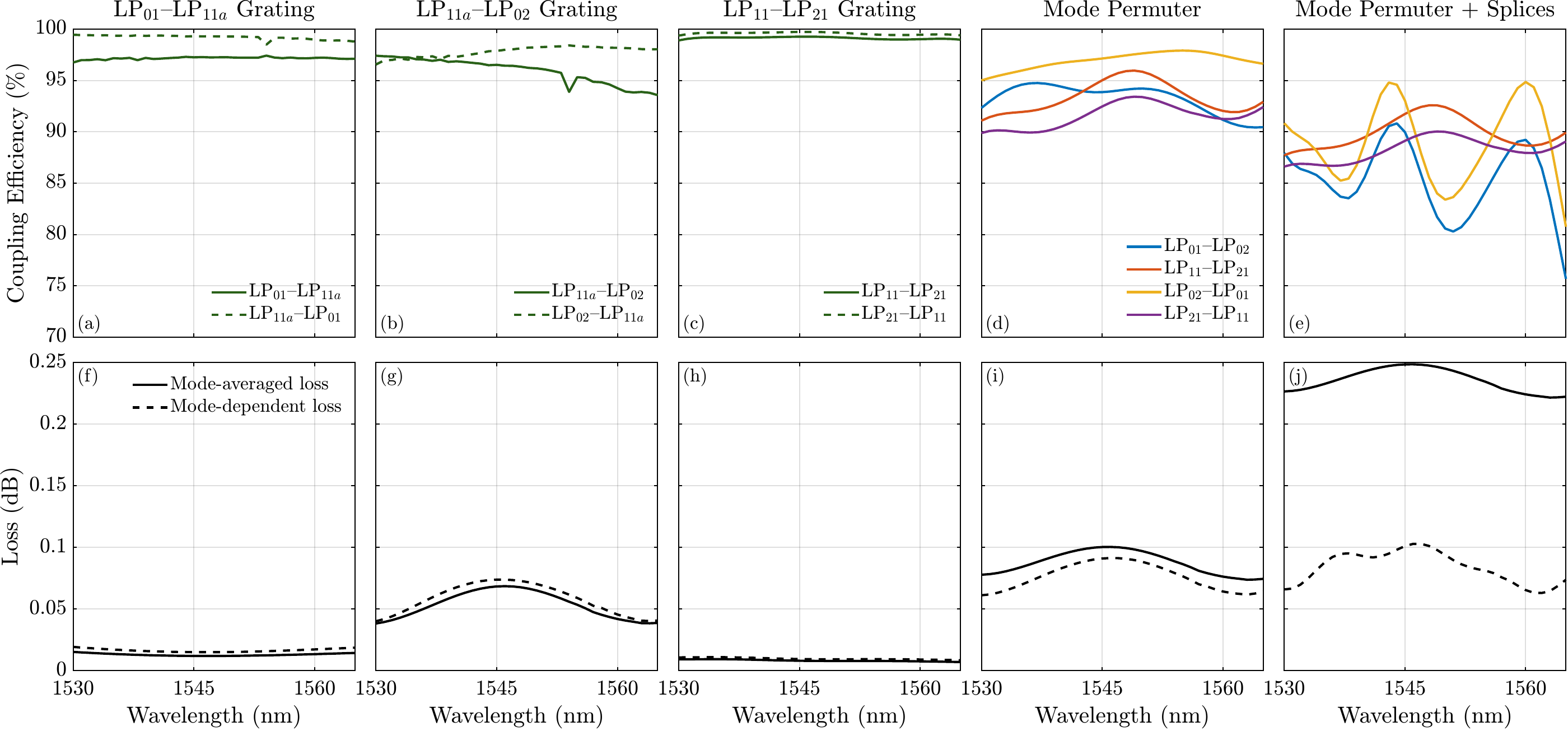}
    \caption{Performance of the individual gratings and the mode permuter designed using the SI transverse index profile in Fig. \ref{fig:step_index_design}. Transmission (a-e) and mode-averaged and mode-dependent losses (f-j) as a function of wavelength over the C-band of the (a, f) $\LP_{01}$\textendash$\LP_{11a}$ grating, (b, g) $\LP_{11a}$\textendash$\LP_{02}$, (c, h) $\LP_{11}$\textendash$\LP_{21}$ grating, (d, i) mode permuter, and (e, j) mode permuter including splicing.}
    \label{fig:step_index_grating_transmission}
\end{figure*}

Fig. \ref{fig:step_index_grating_transmission} shows the performance of the individual gratings, mode permuter, and mode permuter including splicing. 
Fig. \ref{fig:step_index_grating_transmission}(a) and Fig. \ref{fig:step_index_grating_transmission}(f) show the $\LP_{01}$\textendash$\LP_{11a}$ coupling efficiency and MDL STD and MAL, respectively, of the $\LP_{01}$\textendash$\LP_{11a}$ grating.
Coupling efficiency exceeds $97\%$, while the MAL and MDL STD are under $0.03$ dB and $0.02$ dB, respectively.

Fig. \ref{fig:step_index_grating_transmission}(b) and Fig. \ref{fig:step_index_grating_transmission}(g) show the coupling efficiency and losses, respectively, of the $\LP_{11a}$\textendash$\LP_{02}$ grating. 
Coupling efficiency exceeds $94\%$, while the MAL and MDL STD are both under $0.07$ dB.

Fig. \ref{fig:step_index_grating_transmission}(c) and Fig. \ref{fig:step_index_grating_transmission}(h) show the coupling efficiency and losses, respectively, of the $\LP_{11}$\textendash$\LP_{21}$ grating.
The reported coupling efficiencies are an average of the $\LP_{11a}\textrm{\textendash}\LP_{21a}$ and $\LP_{11b}\textrm{\textendash}\LP_{21b}$ coupling efficiencies.
Coupling efficiency exceeds $99\%$, while the MAL and MDL STD are under $0.02$ dB and $0.02$ dB, respectively.

Compared to the $\LP_{01}$\textendash$\LP_{11a}$ and $\LP_{11}$\textendash$\LP_{21}$ gratings, the $\LP_{11a}$\textendash$\LP_{02}$ grating has substantially larger losses. 
This occurs since the $\LP_{11}$\textendash$\LP_{02}$ transition is near two undesired transitions with large coupling coefficients, $\LP_{11}$\textendash$\LP_{21}$ and $\LP_{21}$\textendash$\LP_{31}$, while the $\LP_{01}$\textendash$\LP_{11}$ and $\LP_{11}$\textendash$\LP_{21}$ transitions are closest to the $\LP_{02}$\textendash$\LP_{31}$ transition, which has a very low coupling coefficient. In the next subsection, we use free-form index optimization to address this issue.

Fig. \ref{fig:step_index_grating_transmission}(d) and Fig. \ref{fig:step_index_grating_transmission}(i) show the mode permuter's coupling efficiencies and losses, respectively.
The coupling efficiencies of the $\LP_{01}\textrm{\textendash}\LP_{02}$, $\LP_{11}$\textendash$\LP_{21}$, $\LP_{02}\textrm{\textendash}\LP_{01}$, and $\LP_{21}\textrm{\textendash}\LP_{11}$ transitions exceed $89\%$, while the MAL and MDL STD are under $0.10$ dB and $0.09$ dB, respectively.

Fig. \ref{fig:step_index_grating_transmission}(e) and Fig. \ref{fig:step_index_grating_transmission}(j) show the mode permuter's coupling efficiencies and losses, respectively, when including splicing to and from the transmission fiber.
The coupling efficiencies of the $\LP_{11}$\textendash$\LP_{21}$ and $\LP_{21}$\textendash$\LP_{11}$ couplings exceed $86\%$, while those of the $\LP_{01}$\textendash$\LP_{02}$ and $\LP_{02}$\textendash$\LP_{01}$ couplings only exceed $76\%$. The degradation of the $\LP_{01}$\textendash$\LP_{02}$ mode exchange is from splicing, which causes additional, unwanted coupling between the $\LP_{01}$ and $\LP_{02}$ modes.
At $1565$ nm, efficiencies of the $\LP_{01}$\textendash$\LP_{02}$ and $\LP_{02}$\textendash$\LP_{01}$ couplings decrease by approximately $16\%$.
Additionally, the MAL rises by $0.15$ dB, whereas the MDL shows only a slight increase because the mode permuter compensates for the MDL created by the two splices.

We have shown that a mode permuter with an SI-fiber can perform the desired mode permutations with relatively high coupling efficiency and low loss.
However, splicing loss, undesired coupling between $\LP_{01}$ and $\LP_{02}$ from splicing, and $\LP_{11a}$\textendash$\LP_{02}$ grating loss are key performance limiters. 
In the subsequent design case, we use free-form index optimization to address these issues and design a mode permuter with enhanced performance.

\subsection{Design Case 2: Free-form-Optimized Fiber}
\label{subsec:design_case_optimized}

Following the methodology in Section \ref{sec:mode_perm_design}, we design an improved mode permuter fiber with free-form index optimization. 
The process consists of four phases, each with a different number of iterations $N$, weighting vector $\bm{w}$, step size $\mu$, and set of target propagation constants $\{\boldsymbol{\beta}^{\mathrm{tar}}[l\,m]\}$ over all guided modes $\LP_{lm}$.

\begin{itemize}[itemindent=0.56in]
    \item [\textbf{\textit{Phase 1}}:] Starting with the transmission fiber as the initial state, we perform $N_1$ iterations with 
    \begin{equation*}
        \boldsymbol{\beta}^{\mathrm{tar}} = \boldsymbol{\beta}^{\mathrm{SI}}
    \end{equation*} 
    for all guided modes, where $\boldsymbol{\beta}^{\mathrm{SI}}$ are the propagation constants of the SI-MMF in Fig. \ref{fig:step_index_design}.
    \item [\textbf{\textit{Phase 2}}:]
    We perform $N_2$ iterations where
    \begin{equation*}
        \boldsymbol{\beta}^{\mathrm{tar}}[02] = \boldsymbol{\beta}^{\mathrm{SI}}[02] - \delta.
    \end{equation*}
    and the rest of the $\boldsymbol{\beta}^{\mathrm{tar}}$ are identical to $\boldsymbol{\beta}^{\mathrm{SI}}$. In the SI fiber, the $\LP_{11}\textrm{\textendash}\LP_{02}$ transition is close to the transitions adjacent to it on each side, causing some undesired coupling. Phase 2 rectifies this in the optimized design by shifting the $\LP_{02}$ mode downwards and thus reordering the spacings to create more separation from the $\LP_{11}\textrm{\textendash}\LP_{02}$ transition. This slight modification increases the separation between the $\LP_{11}\textrm{\textendash}\LP_{02}$ transition and all adjacent transitions by $50\%$, substantially reducing parasitic coupling.
    \item [\textbf{\textit{Phase 3}}:]
    We perform $N_3$ iterations where
    \begin{align*}
        &\boldsymbol{\beta}^{\mathrm{tar}}[11] = \boldsymbol{\beta}^{\mathrm{SI}}[11] + \delta, \\
        &\boldsymbol{\beta}^{\mathrm{tar}}[21] = \boldsymbol{\beta}^{\mathrm{SI}}[21] + \delta.
    \end{align*}
    and the rest of the $\boldsymbol{\beta}^{\mathrm{tar}}$ are identical to $\boldsymbol{\beta}^{\mathrm{SI}}$. This shifts $\LP_{11}$ and $\LP_{21}$ upwards, further separating all couplings and isolating the desired transitions.
    \item [\textbf{\textit{Phase 4}}:]
    The last stage of fine-tuning consists of $N_4$ iterations with the $\{\boldsymbol{\beta}^{\mathrm{tar}}\}$ identical to those in Phase 3, but with the weightings changed to target undesired transitions that continue to interfere with the ideal permutation scheme. More iterations would isolate the desired transitions even further, but the additional RI perturbations cause more splicing loss. We select the value of $n_4$ to reflect the best compromise. 

\end{itemize}
After sweeping each hyperparameter, we set $\delta = 6k_0 \times 10^{-4}$, $N_1 = 250$, $N_2 = 40$, $N_3 = 25$, and $N_4 = 15$.
The step size $\mu$ is $5 \times 10^{-6}$ for Phases 1-3 and is halved in Phase 4. The weighting $w$ is increased by a factor of 4 for the particular mode or modes targeted in each phase and by a factor of 2 for the guided non-signal modes.

\begin{figure*}
    \centering
    \includegraphics[width=1\linewidth]{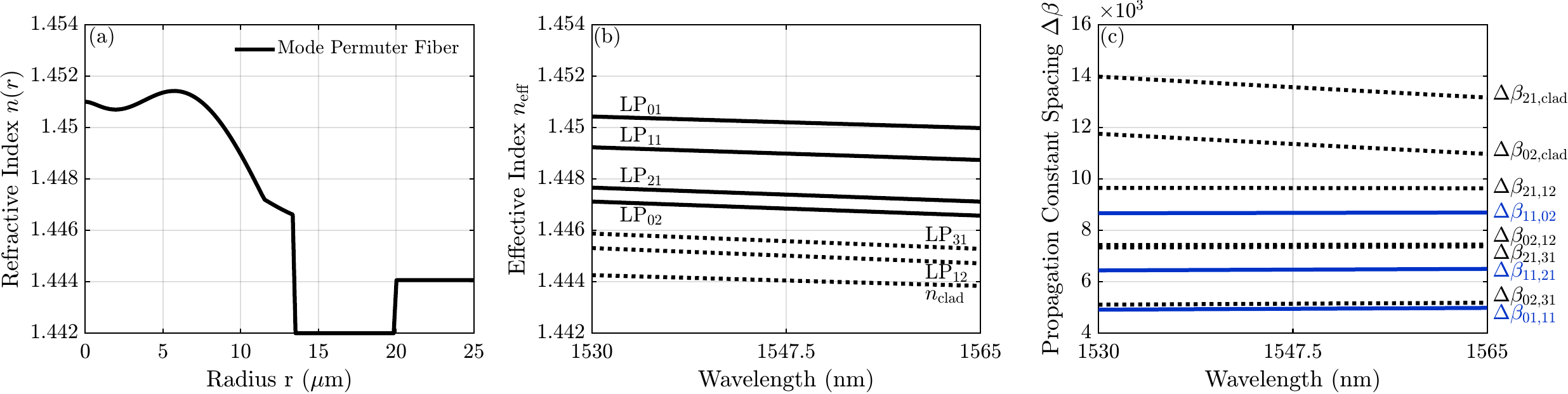}
    \caption{Mode permuter fiber designed using free-form index optimization: (a) Transverse index profile, (b) guided mode effective indices and cladding index as a function of wavelength, and (c) select propagation constant spacings as a function of wavelength. In (b), guided modes not used for signal propagation and the cladding index are dashed. In (c), propagation constant spacings for the $\LP_{01}\textrm{\textendash}\LP_{11}$, $\LP_{02}\textrm{\textendash}\LP_{12}$, and $\LP_{11}\textrm{\textendash}\LP_{21}$ transitions are in blue, while all others are dashed.}
    \label{fig:optimized_index_design}
\end{figure*}

Fig. \ref{fig:optimized_index_design} shows the transverse index profile, effective indices, and propagation constant spacings of the new design obtained with free-form index optimization.
The splicing MAL and MDL STD are smaller compared to the SI design, assuming values under $0.03$ dB and $0.02$ dB, respectively. 
The propagation constant spacings $\Delta\beta_{01,11}$, $\Delta\beta_{11,02}$, and $\Delta\beta_{11,21}$ vary over wavelength by $67.1$ $\mathrm{m}^{-1}$, $26.7$ $\mathrm{m}^{-1}$, and $57.5$ $\mathrm{m}^{-1}$, respectively.
The separation between the $\LP_{01}$\textendash$\LP_{11}$, $\LP_{11}$\textendash$\LP_{02}$, and $\LP_{11}$\textendash$\LP_{21}$ transitions and adjacent transitions is greater than $875$ $\mathrm{m}^{-1}$, excluding the inefficient $\LP_{02}\textrm{\textendash}\LP_{31}$ transition.
The beat lengths of the $\TE_{01}/\HE_{21}$ and $\TM_{01}/\HE_{21}$ modes increase to about $0.78$ $\mathrm{m}$ and $5.7$ $\mathrm{m}$, respectively.

Comparing Fig. \ref{fig:optimized_index_design}(a) and Fig. \ref{fig:step_index_design}(a) with Fig. \ref{fig:transmission_fiber}(a), the free-form-optimized fiber clearly more closely resembles the transmission fiber than the SI fiber does, explaining the low splicing loss.
Moreover, we are able to achieve this while further isolating the desired transitions. The greater consistency of the propagation constant spacings across all wavelengths and the longer beat lengths of $LP_{11}$ improve the coupling efficiency as well. 

{\renewcommand{\arraystretch}{1.2}
\setlength{\tabcolsep}{2pt}
\begin{table}
\centering

\begin{threeparttable}
\caption{Grating Parameters for Free-form-optimized Fiber}
\begin{tabular}{c|c|c|c}
\toprule
Grating & $\LP_{01}$\textendash$\LP_{11a}$  & $\LP_{11a}$\textendash$\LP_{02}$ & $\LP_{11}$\textendash$\LP_{21}$ \\ \hline 
{UV Illumination Type} & Single & Single & Double \\
{UV Illumination Angle(s) $\psi$} & $0$ & $0$ & ${\pi}/{6}$ and $-{\pi}/{6}$ \\
{Modulation Depth $\chi$} & $20\times 10^{-5}$ & $30\times 10^{-5}$ & {$10\times 10^{-5}$}\\
{Grating period $\Lambda$ ($\mu$m)} & $1234.1$ & $706.6$ & {$954.3$}\\
{Grating length $L$ (cm)} & $6.05$ & $5.48$ & {$8.02$}\\
{Chirp parameter $\alpha$} & 4 & 4 & 4 \\
\bottomrule
\end{tabular}
\label{table:freeform_grating_params}
\end{threeparttable}
\end{table}
}

Table \ref{table:freeform_grating_params} provides the UV illumination type and values of the modulation depth $\chi$, grating period $\Lambda$, grating length $L$, and chirp $\alpha$ for the $\LP_{01}$\textendash$\LP_{11a}$, $\LP_{11a}$\textendash$\LP_{02}$, and $\LP_{11}\textrm{\textendash}\LP_{21}$ gratings designed using the free-form-optimized fiber. 
We obtain the grating parameters following the same methodology as in the step-index design case.

\begin{figure*}
    \centering
    \includegraphics[width=1\linewidth]{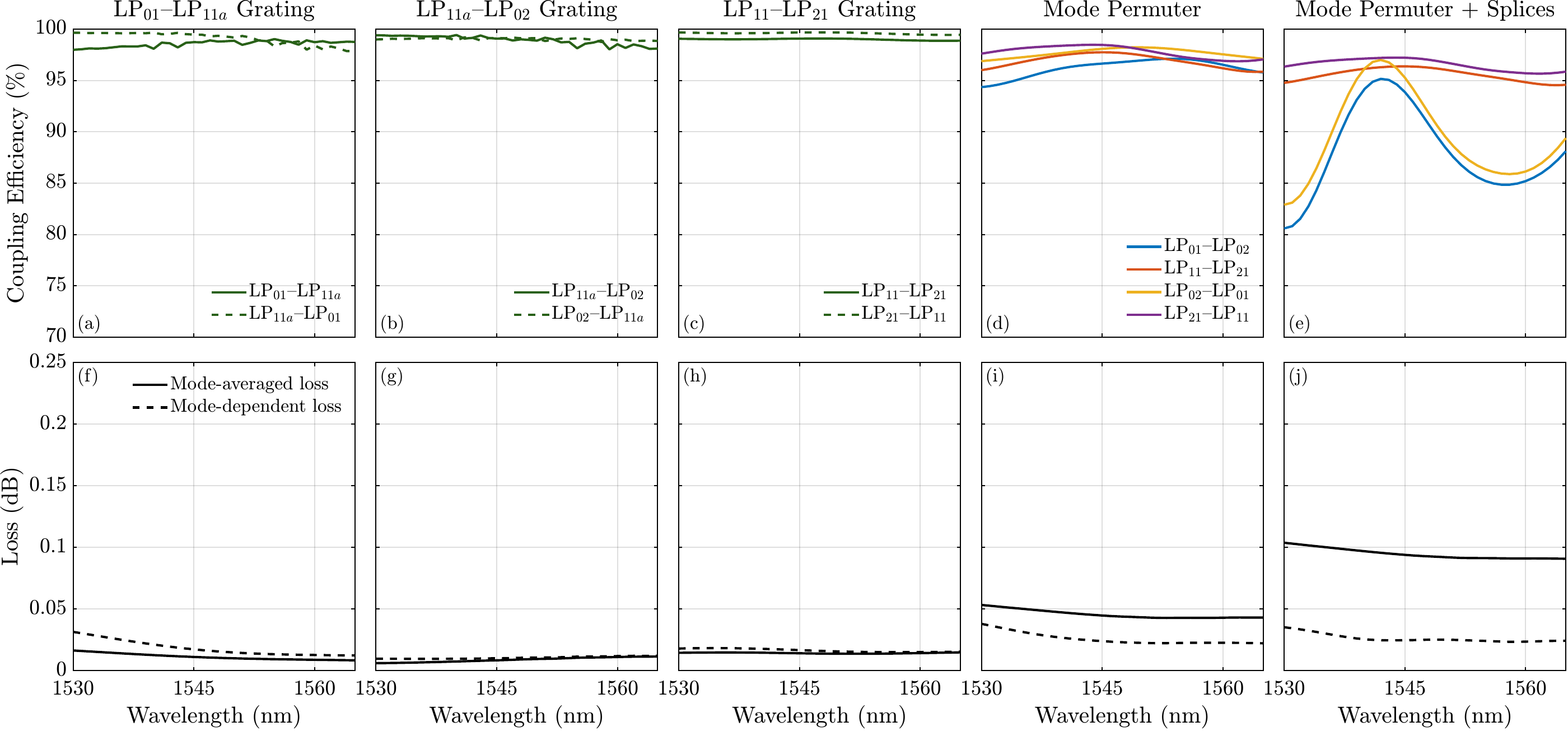}
    \caption{Performance of individual gratings and mode permuter designed using the free-form-optimized transverse index profile in Fig. \ref{fig:optimized_index_design}. Coupling efficiencies (a-e) and mode-averaged and mode-dependent losses (f-j) as a function of wavelength over the C-band of the (a, f) $\LP_{01}$\textendash$\LP_{11a}$ grating, (b, g) $\LP_{11a}$\textendash$\LP_{02}$, (c, h) $\LP_{11}\textrm{\textendash}\LP_{21}$ grating, (d, i) mode permuter, and (e, j) mode permuter including splicing.}
    \label{fig:freeform_opt_coupling_transmission}
\end{figure*}

Fig. \ref{fig:freeform_opt_coupling_transmission} shows the performance of the individual gratings, mode permuter, and mode permuter including splicing. 
Fig. \ref{fig:freeform_opt_coupling_transmission}(a) and Fig. \ref{fig:freeform_opt_coupling_transmission}(f) show the coupling efficiency and losses, respectively, of the $\LP_{01}$\textendash$\LP_{11a}$ grating.
Coupling efficiency exceeds $97\%$, while the MAL and MDL STD are under $0.04$ dB and $0.02$ dB, respectively.
Compared to the SI-fiber grating, there is slightly greater loss owing to the $\LP_{02}\textrm{\textendash}\LP_{31}$ transition being closer to the $\LP_{01}\textrm{\textendash}\LP_{11}$ transition.

Fig. \ref{fig:freeform_opt_coupling_transmission}(b) and Fig. \ref{fig:freeform_opt_coupling_transmission}(g) show the coupling efficiency and losses, respectively, for the $\LP_{11a}$\textendash$\LP_{02}$ grating. 
Coupling efficiency exceeds $98\%$, while the MAL and MDL STD are both under $0.02$ dB.
Compared to the SI-fiber grating, the losses are over $0.05$ dB less at $1545$ nm, validating our free-form index optimization.

Fig. \ref{fig:freeform_opt_coupling_transmission}(c) and Fig. \ref{fig:freeform_opt_coupling_transmission}(h) show the coupling efficiencies and losses, respectively, over the C-band for the $\LP_{11}\textrm{\textendash}\LP_{21}$ grating.
The reported coupling efficiencies are an average of those of the $\LP_{11a}\textrm{\textendash}\LP_{21a}$ and $\LP_{11b}\textrm{\textendash}\LP_{21b}$ couplings.
Coupling efficiency exceeds $99\%$, while the MAL and MDL STD are both under $0.02$ dB.

Fig. \ref{fig:freeform_opt_coupling_transmission}(d) and Fig. \ref{fig:freeform_opt_coupling_transmission}(i) show the mode permuter's coupling efficiencies and losses, respectively.
The efficiencies of the $\LP_{01}\textrm{\textendash}\LP_{02}$, $\LP_{11}\textrm{\textendash}\LP_{21}$, $\LP_{02}\textrm{\textendash}\LP_{01}$, and $\LP_{21}\textrm{\textendash}\LP_{11}$ couplings exceed $94\%$, while the MAL and MDL STD are under $0.06$ dB and $0.04$ dB, respectively.

Fig. \ref{fig:freeform_opt_coupling_transmission}(e) and Fig. \ref{fig:freeform_opt_coupling_transmission}(j) show the mode permuter's coupling efficiencies and losses, respectively, when including splicing to and from the transmission fiber.
The efficiencies of the $\LP_{11}$\textendash$\LP_{21}$ and $\LP_{21}$\textendash$\LP_{11}$ couplings exceed $95\%$, while those of the $\LP_{01}$\textendash$\LP_{02}$ and $\LP_{02}$\textendash$\LP_{01}$ couplings exceed $81\%$.
The MAL rises by about $0.05$ dB, whereas the MDL shows only a slight increase, similar to the SI design. 

A comparison of Fig. \ref{fig:freeform_opt_coupling_transmission} with Fig. \ref{fig:step_index_grating_transmission} demonstrates the efficacy of our index optimization strategy in enhancing performance.
By leveraging the fine control over fiber propagation constants that index optimization provides, we achieve a substantial reduction in loss while increasing coupling efficiencies for all desired mode exchanges. 
Comparing Fig. \ref{fig:freeform_opt_coupling_transmission}(i) with Fig. \ref{fig:freeform_opt_coupling_transmission}(j) reveals that, despite the new design's reduced splicing loss, splicing to and from the mode permuter contributes more loss than the device itself.
Since two splices are required for every mode permuter, this issue warrants further study.

\subsection{Link Performance}
\label{subsec:system_performance}
\begin{figure}
    \centering
    \includegraphics[width=0.7\linewidth]{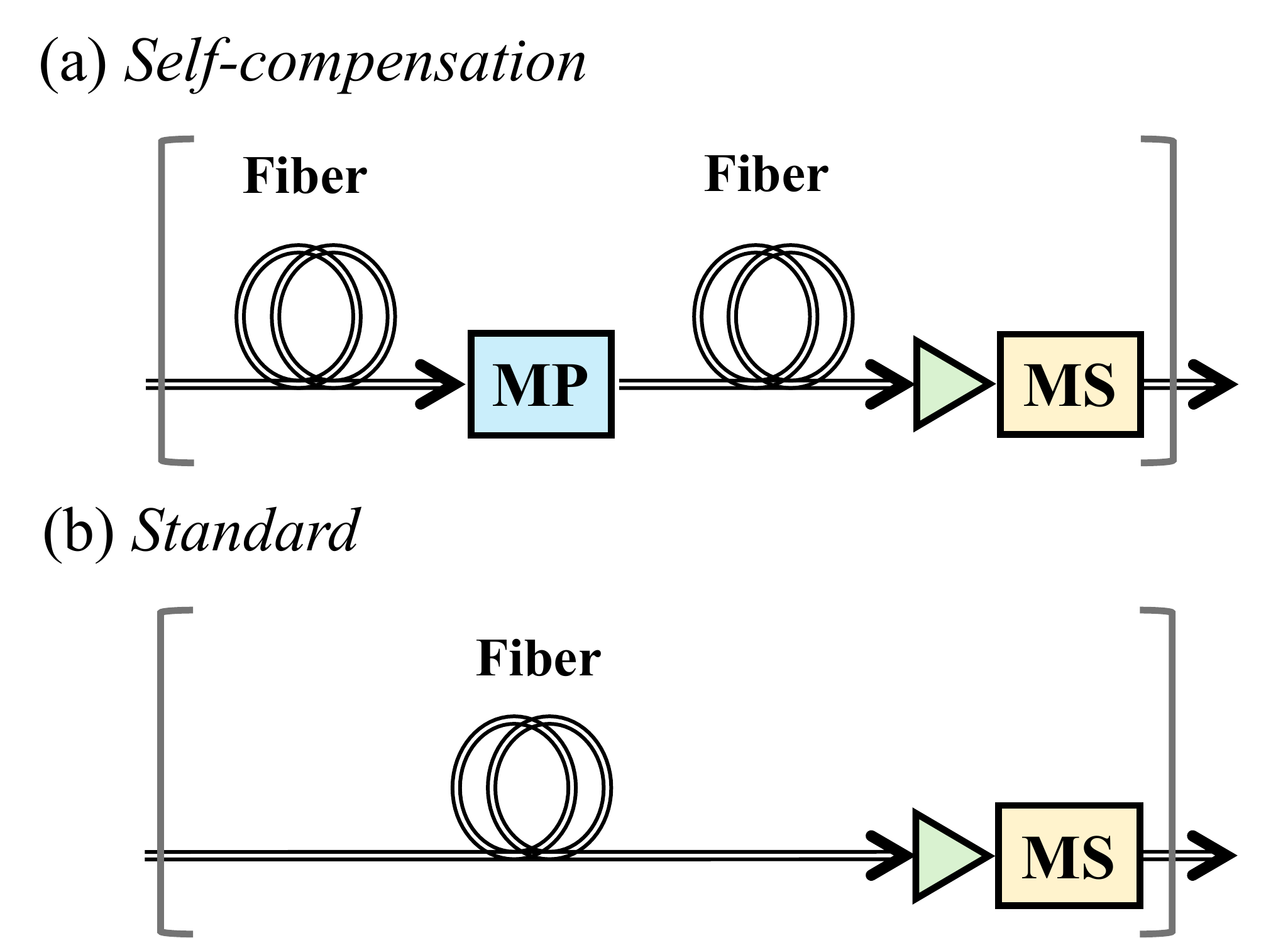}
    \caption{Diagrams of one span of a long-haul MDM transmission link with periodic amplification and mode scrambling at the end of every span. (a) Self-compensation scheme using a mode permuter in the middle of each span. (b) Standard scheme employing only mode scramblers, serving as a baseline. MP: Mode Permuter, MS: Mode Scrambler.}
    \label{fig:system_diagrams}
\end{figure}

We consider two long-haul MDM link architectures to evaluate the system performance of our mode permuter device, as shown in Fig. \ref{fig:system_diagrams}. 
Fig. \ref{fig:system_diagrams}(a) illustrates the self-compensation architecture, which incorporates a mode permuter in the middle of each span, as well as amplification and mode scrambling at the end of each span.
Fig. \ref{fig:system_diagrams}(b) illustrates the standard architecture, which omits the mode permuter, serving as a baseline for the GD STD.

We perform numerical multisection simulations to calculate the GD STD of each link in Fig. \ref{fig:system_diagrams}. 
For the self-compensation architecture, we use the free-form-optimized mode permuter with characteristics given in Fig. \ref{fig:freeform_opt_coupling_transmission}.
To highlight the harmful effect of splicing on GD STD reduction, we simulate the link including and excluding impairments from splicing between the mode permuter and transmission fiber. 
For both the self-compensation and standard architectures, the mode scrambler is assumed to equally distribute power between all modes, and the amplifier is assumed to have no MDL.
We select a span length $\lenS=50~\textrm{km}$ and assume strong random intra-group coupling in the transmission fiber \cite{vijay_modal_2024}.
We consider three random inter-group coupling lengths $\Linter$ of $500~\textrm{km}$, $2000~\textrm{km}$, and $\infty~\textrm{km}$. 
$\Linter=\infty~\textrm{km}$ represents a scenario with no random inter-group coupling, in which self-compensation performs the best.
 
\begin{figure*}
    \centering
    \includegraphics[width=1\linewidth]{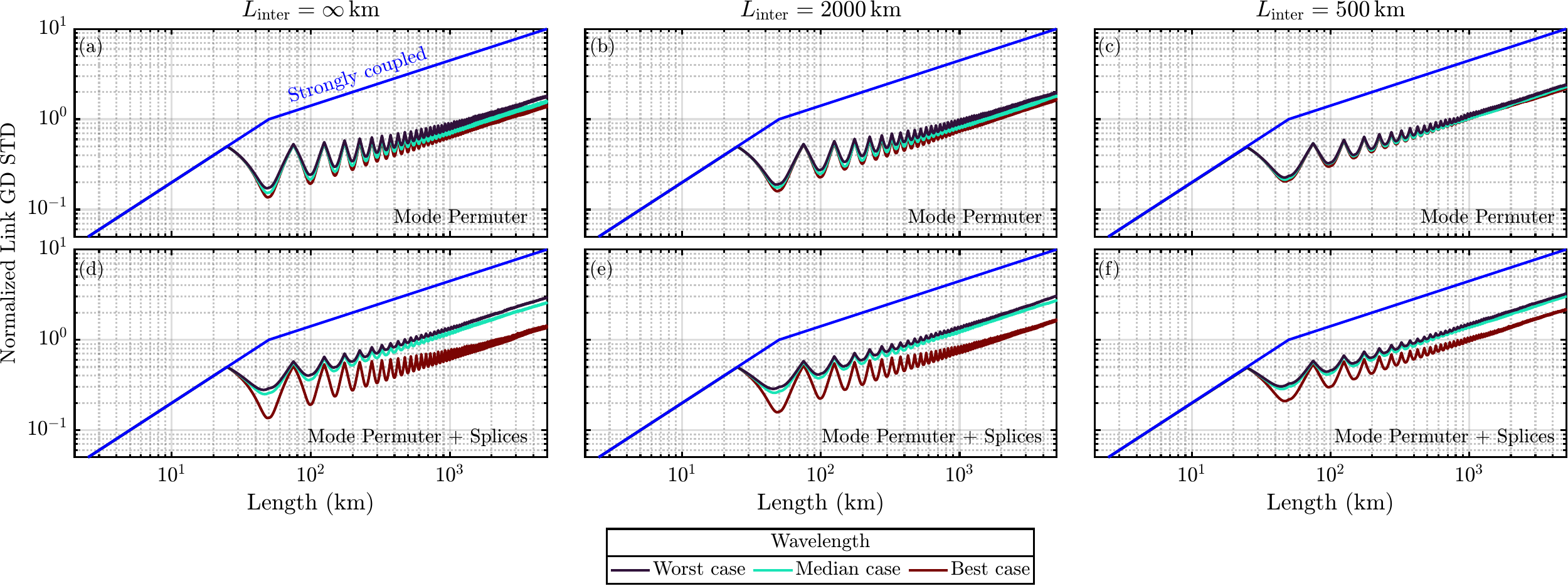}
    \caption{Simulation of self-compensation. Normalized link GD STD as a function of length for worst-, median-, and best-performing wavelengths when (a, d) $\Linter=\infty$ km, (b, e) $\Linter=2000$ km, and (c, f) $\Linter=500$ km using the transfer matrix of the free-form-optimized mode permuter (a-c) excluding splicing and (d-f) including splicing.}
    \label{fig:Linters_varied}
\end{figure*}

Fig.~\ref{fig:Linters_varied}(a)-(c) shows the normalized GD STD as a function of link length over $K=100$ spans for the worst-, median-, and best-performing wavelength channels in the C-band, excluding splicing.
The dashed blue line indicates the GD STD of a link employing the standard architecture in Fig. \ref{fig:system_diagrams}(b), which has GD STD proportional to the square root of the number of spans.
The rise and fall of GD STD in each span clearly illustrate how effective the mode permuter is at compensating GD.
The self-compensating link significantly surpasses the baseline link for all inter-group coupling lengths and wavelengths, exhibiting an end-to-end GD STD that is approximately 4 times smaller after $5000~\textrm{km}$.
Due to the stable coupling efficiency over wavelength, only minor differences in GD STD are observed between the best- and worst-performing wavelengths for $\Linter=\infty$ and $\Linter=2000$ km. 
These differences nearly vanish for $\Linter=500$ km because the power transfer between modes resulting from random inter-group coupling is comparable to the non-ideality of the mode permuter power transfer. 

Fig.~\ref{fig:Linters_varied}(d)-(f) shows the normalized GD STD as a function of link length over $K=100$ spans for the worst-, median-, and best-case wavelength channels in the C-band including splicing.
While the self-compensating link continues to outperform the baseline link, the disparity between the best and worst-performing wavelengths is much more noticeable due to the effect of splicing.
The best- and worst-performing wavelengths differ by a factor of $1.5$ and $1.9$ for random inter-group coupling lengths $\Linter=500$ and $\Linter=\infty$ km, respectively.

\begin{figure}
    \centering
    \includegraphics[width=1\linewidth]{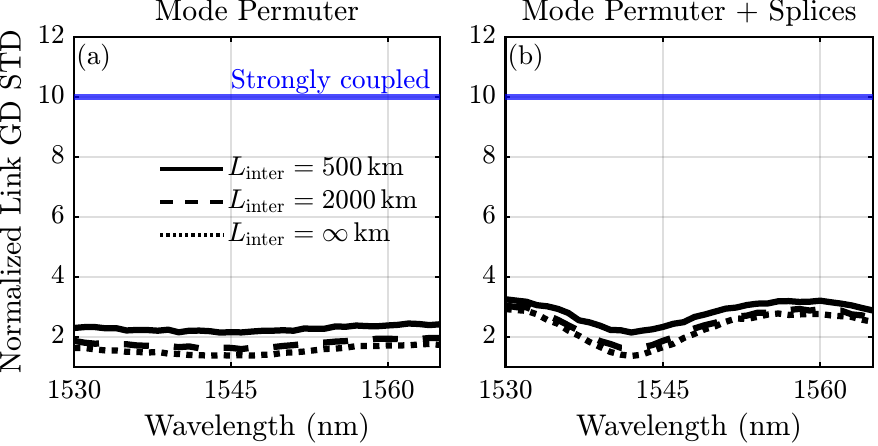}
    \caption{Normalized link GD STD as a function of wavelength after propagation over 5000 km for $\Linter$ equal to $\infty$, $2000$, and $500$ km using the transfer matrix of the free-form-optimized mode permuter (a) excluding splicing and (b) including splicing.}
    \label{fig:Linters_varied_allwl}
\end{figure}

Fig.~\ref{fig:Linters_varied_allwl}(a) shows the normalized GD STD as a function of wavelength after $K=100$ spans at various random inter-group coupling lengths excluding splicing. 
The dashed blue line indicates the link GD STD after $K=100$ spans of the reference link, with GD STD proportional to the square root of the number of spans.
The GD STD varies little over the C-band, tracking well with the stable coupling efficiencies in Fig. \ref{fig:freeform_opt_coupling_transmission}(d).
We clearly see that random inter-group coupling limits GD compensation.
The GD STD increases by $50\%$ when the random inter-group coupling length decreases from $\Linter=\infty$ km to $\Linter=500$ km.

Fig.~\ref{fig:Linters_varied_allwl}(b) shows the normalized GD STD as a function of wavelength after $K=100$ spans at various random inter-group coupling lengths including splicing. 
The GD STD follows the opposite trend of the $\LP_{01}\textrm{\textendash}\LP_{02}$ and $\LP_{02}\textrm{\textendash}\LP_{01}$ lines in Fig. \ref{fig:freeform_opt_coupling_transmission}(e).
At $1530$ nm, the GD STD is largest because the $\LP_{01}$ and $\LP_{02}$ coupling is most degraded due to splicing at this wavelength.
Similarly, at $1540$ nm, the GD STD is the smallest since the $\LP_{01}$ and $\LP_{02}$ coupling is least affected at this wavelength.

Fig.~\ref{fig:Linters_varied} and Fig.~\ref{fig:Linters_varied_allwl} clearly demonstrate that a well-designed mode permuter is highly effective in reducing link GD STD, decreasing GD STD by more than a factor of $3.13$ across the C-band. If a mode scrambler were used in its place, the GD STD would decrease by only a factor of $\sqrt{2}\approx1.41$ \cite{vijay_modal_2024}. 

Comparing Fig.~\ref{fig:Linters_varied}(a)-(c) with Fig.~\ref{fig:Linters_varied}(d)-(f) shows the significant effect of unwanted coupling between $\LP_{01}$ and $\LP_{02}$ during splicing. 
While self-compensation offers a significant reduction in GD STD, it is crucial to minimize any unwanted coupling, particularly from splicing or from other link components.

\section{Discussion}
\label{sec:discussion}

\subsection{Splicing}
As noted in Section \ref{sec:results}, undesired  coupling between the $\LP_{01}$ and $\LP_{02}$ modes and loss due to splicing remain the primary performance impairments even for the free-form-optimized design.
These issues can potentially be ameliorated by optimizing the fusion splicing process or using intermediate bridge fibers.
Reference \cite{scarnera_splice_2024} shows how varying the arc length duration during fusion splicing reduces crosstalk between $\LP_{01}$ and $\LP_{02}$ from $-15$ dB to $-25$ dB for an SMF to MMF splice.
This would significantly minimize the wavelength variation of the link GD STD. 
Hence, optimizing the splicing process may be crucial for self-compensation and warrants further study.

\subsection{Extending to More Mode Groups}
The principles used to determine an ideal mode permutation scheme can be extended to any number of mode groups $N_g$ using the theoretical framework described in Section \ref{sec:mode_permutation_theory}. \cite{vijay_closed-form_2025} describes a general approach for arbitrary $N_g$. While such mode permutation schemes exist, their implementation becomes more difficult as $N_g$ increases and more challenges arise.

A system with more mode groups requires the permuter to perform more mode exchanges. Some of these exchanges may occur between mode groups with vastly different propagation constants, and coupling them directly could cause coupling into the cladding and subsequent loss. Thus, such distant mode exchanges would need to be performed stepwise, sequentially exchanging powers between adjacent mode groups until the final destination is reached. This increases the number of gratings required.

With more modes, there are more potential paths for undesired coupling. Every propagation constant spacing for an undesired transition must be separated from the desired transitions, and both types become more numerous as $N_g$ rises, further constraining the design space. At some point, separating all these spacings within a single fiber may become intractable, necessitating multiple mode permutation fibers spliced together. This will impact the splicing loss.

The mode permuter device must be designed jointly with the transmission fiber, which has its own unique design challenges when $N_g$ is large \cite{vijay_closed-form_2025}. These are topics for further study.

\section{Conclusion}
\label{sec:conclusion}

We described a self-compensation scheme for an MDM link supporting $D=12$ spatial and polarization modes. 
This scheme relied on the modes in the first two mode groups having equal and opposite modal dispersions compared to those of the third mode group, along with a mode permuter that exchanges all power between the third mode group and the first two mode groups. 
We proposed a mode permuter design with a cascade of four LPFGs exchanging the following pairs of modes: $\LP_{01}$\textendash$\LP_{11a}$, $\LP_{11a}$\textendash$\LP_{02}$, $\LP_{01}$\textendash$\LP_{11a}$, and $\LP_{11}\textrm{\textendash}\LP_{21}$.
Then, we described a strategy for designing the mode permuter fiber transverse index profile and the grating index profiles.

We first designed a mode permuter with gratings inscribed on SI fiber, using grid search to find an SI fiber with low splicing loss to the transmission fiber and well-separated propagation constant spacings for the $\LP_{01}\textrm{\textendash}\LP_{11}$, $\LP_{11}\textrm{\textendash}\LP_{02}$, and $\LP_{11}\textrm{\textendash}\LP_{02}$ transitions.
Then, chirped gratings were designed for each mode exchange to have low MAL and MDL, while maintaining high transmission over the C-band.
This resulted in a mode permuter design with transmission over $75\%$ for all desired mode power exchanges, a MAL of less than $0.25$ dB, and an MDL of less than $0.11$ dB over the C-band.
From analyzing this design, we found that splicing loss, undesired coupling between $\LP_{01}$ and $\LP_{02}$ from splicing, and $\LP_{11a}$\textendash$\LP_{02}$ grating loss are key performance limiters.

To address these issues, we designed an improved transverse index profile using free-form index optimization. 
This yielded a mode permuter design with transmission over $81\%$ for all desired mode exchanges, a MAL of less than $0.11$ dB, and an MDL of less than $0.04$ dB over the C-band.

We numerically evaluated the designs through link simulations and investigated the reduction in GD spread for different levels of random inter-group coupling in the fiber. 
Our results showed that periodic mode permutation and mode scrambling reduces a link's GD STD by a factor exceeding $3.13$ compared to a link relying solely on periodic mode scrambling.

\section*{Acknowledgment}
This project was supported by Ciena Corporation, Stanford Shoucheng Zhang Graduate Fellowship, and Stanford Graduate Fellowship.
Much of the computing for this project was performed on the Sherlock cluster at Stanford University. 
We thank the Stanford Research Computing Center for providing this cluster and technical support.

\bibliographystyle{IEEEtran}
\bibliography{references}

\begin{thebibliography}{10}
\providecommand{\url}[1]{#1}
\csname url@samestyle\endcsname
\providecommand{\newblock}{\relax}
\providecommand{\bibinfo}[2]{#2}
\providecommand{\BIBentrySTDinterwordspacing}{\spaceskip=0pt\relax}
\providecommand{\BIBentryALTinterwordstretchfactor}{4}
\providecommand{\BIBentryALTinterwordspacing}{\spaceskip=\fontdimen2\font plus
\BIBentryALTinterwordstretchfactor\fontdimen3\font minus \fontdimen4\font\relax}
\providecommand{\BIBforeignlanguage}[2]{{%
\expandafter\ifx\csname l@#1\endcsname\relax
\typeout{** WARNING: IEEEtran.bst: No hyphenation pattern has been}%
\typeout{** loaded for the language `#1'. Using the pattern for}%
\typeout{** the default language instead.}%
\else
\language=\csname l@#1\endcsname
\fi
#2}}
\providecommand{\BIBdecl}{\relax}
\BIBdecl

\bibitem{srinivas_modeling_2021}
\BIBentryALTinterwordspacing
H.~Srinivas, J.~D. Downie, J.~Hurley, X.~Liang, J.~Himmelreich, J.~K. Perin, D.~A.~A. Mello, and J.~M. Kahn, ``Modeling and {Experimental} {Measurement} of {Power} {Efficiency} for {Power}-{Limited} {SDM} {Submarine} {Transmission} {Systems},'' \emph{Journal of Lightwave Technology}, vol.~39, no.~8, pp. 2376--2386, Apr. 2021, conference Name: Journal of Lightwave Technology. [Online]. Available: \url{https://ieeexplore.ieee.org/document/9314868}
\BIBentrySTDinterwordspacing

\bibitem{sinkin_maximum_2017}
\BIBentryALTinterwordspacing
O.~V. Sinkin, A.~V. Turukhin, W.~W. Patterson, M.~A. Bolshtyansky, D.~G. Foursa, and A.~N. Pilipetskii, ``Maximum {Optical} {Power} {Efficiency} in {SDM}-{Based} {Optical} {Communication} {Systems},'' \emph{IEEE Photonics Technology Letters}, vol.~29, no.~13, pp. 1075--1077, Jul. 2017, conference Name: IEEE Photonics Technology Letters. [Online]. Available: \url{https://ieeexplore.ieee.org/document/7919195}
\BIBentrySTDinterwordspacing

\bibitem{sinkin_sdm_2018}
O.~V. Sinkin, A.~V. Turukhin, Y.~Sun, H.~G. Batshon, M.~V. Mazurczyk, C.~R. Davidson, J.-X. Cai, W.~W. Patterson, M.~A. Bolshtyansky, D.~G. Foursa, and A.~N. Pilipetskii, ``{SDM} for {Power}-{Efficient} {Undersea} {Transmission},'' \emph{Journal of Lightwave Technology}, vol.~36, no.~2, pp. 361--371, Jan. 2018, conference Name: Journal of Lightwave Technology.

\bibitem{cai_9_2022}
\BIBentryALTinterwordspacing
J.-X. Cai, G.~Vedala, Y.~Hu, O.~V. Sinkin, M.~A. Bolshtyansky, D.~G. Foursa, and A.~N. Pilipetskii, ``9 {Tb}/s {Transmission} {Using} 29 {mW} {Optical} {Pump} {Power} {Per} {EDFA} {With} 1.24 {Tb}/s/{W} {Optical} {Power} {Efficiency} {Over} 15,050 km,'' \emph{Journal of Lightwave Technology}, vol.~40, no.~6, pp. 1650--1657, Mar. 2022, conference Name: Journal of Lightwave Technology. [Online]. Available: \url{https://ieeexplore.ieee.org/document/9626578}
\BIBentrySTDinterwordspacing

\bibitem{klaus_advanced_2017}
\BIBentryALTinterwordspacing
W.~Klaus, B.~J. Puttnam, R.~S. Luis, J.~Sakaguchi, J.-M.~D. Mendinueta, Y.~Awaji, and N.~Wada, ``Advanced space division multiplexing technologies for optical networks [{Invited}],'' \emph{Journal of Optical Communications and Networking}, vol.~9, no.~4, pp. C1--C11, Apr. 2017, conference Name: Journal of Optical Communications and Networking. [Online]. Available: \url{https://ieeexplore.ieee.org/document/7901441}
\BIBentrySTDinterwordspacing

\bibitem{winzer_chapter_2013}
\BIBentryALTinterwordspacing
P.~J. Winzer, R.~Ryf, and S.~Randel, ``Chapter 10 - {Spatial} {Multiplexing} {Using} {Multiple}-{Input} {Multiple}-{Output} {Signal} {Processing},'' in \emph{Optical {Fiber} {Telecommunications} ({Sixth} {Edition})}, ser. Optics and {Photonics}, I.~P. Kaminow, T.~Li, and A.~E. Willner, Eds.\hskip 1em plus 0.5em minus 0.4em\relax Boston: Academic Press, Jan. 2013, pp. 433--490. [Online]. Available: \url{https://www.sciencedirect.com/science/article/pii/B9780123969606000109}
\BIBentrySTDinterwordspacing

\bibitem{srinivas_efficient_2023}
\BIBentryALTinterwordspacing
H.~Srinivas, O.~Krutko, and J.~M. Kahn, ``Efficient {Integrated} {Multimode} {Amplifiers} for {Scalable} {Long}-{Haul} {SDM} {Transmission},'' \emph{Journal of Lightwave Technology}, vol.~41, no.~15, pp. 4989--5002, Aug. 2023, conference Name: Journal of Lightwave Technology. [Online]. Available: \url{https://ieeexplore.ieee.org/document/10064021}
\BIBentrySTDinterwordspacing

\bibitem{jensen_demonstration_2015}
\BIBentryALTinterwordspacing
R.~V. Jensen, L.~Grüner-Nielsen, N.~H.~L. Wong, Y.~Sun, Y.~Jung, and D.~J. Richardson, ``\BIBforeignlanguage{EN}{Demonstration of a 9 {LP}-{Mode} {Transmission} {Fiber} with {Low} {DMD} and {Loss}},'' in \emph{\BIBforeignlanguage{EN}{Optical {Fiber} {Communication} {Conference} (2015), paper {W2A}.34}}.\hskip 1em plus 0.5em minus 0.4em\relax Optica Publishing Group, Mar. 2015, p. W2A.34. [Online]. Available: \url{https://opg.optica.org/abstract.cfm?uri=OFC-2015-W2A.34}
\BIBentrySTDinterwordspacing

\bibitem{ryf_mode-multiplexed_2015}
\BIBentryALTinterwordspacing
R.~Ryf, N.~K. Fontaine, H.~Chen, B.~Guan, B.~Huang, M.~Esmaeelpour, A.~H. Gnauck, S.~Randel, S.~J.~B. Yoo, A.~M.~J. Koonen, R.~Shubochkin, Y.~Sun, and R.~Lingle, ``\BIBforeignlanguage{EN}{Mode-multiplexed transmission over conventional graded-index multimode fibers},'' \emph{\BIBforeignlanguage{EN}{Optics Express}}, vol.~23, no.~1, pp. 235--246, Jan. 2015, publisher: Optica Publishing Group. [Online]. Available: \url{https://opg.optica.org/oe/abstract.cfm?uri=oe-23-1-235}
\BIBentrySTDinterwordspacing

\bibitem{fontaine_characterization_2013}
\BIBentryALTinterwordspacing
N.~K. Fontaine, R.~Ryf, M.~A. Mestre, B.~Guan, X.~Palou, S.~Randel, Y.~Sun, L.~Grüner-Nielsen, R.~V. Jensen, and R.~Lingle, ``Characterization of space-division multiplexing systems using a swept-wavelength interferometer,'' in \emph{2013 {Optical} {Fiber} {Communication} {Conference} and {Exposition} and the {National} {Fiber} {Optic} {Engineers} {Conference} ({OFC}/{NFOEC})}, Mar. 2013, pp. 1--3. [Online]. Available: \url{https://ieeexplore.ieee.org/abstract/document/6533121}
\BIBentrySTDinterwordspacing

\bibitem{krutko_ultra-low-loss_2025}
\BIBentryALTinterwordspacing
O.~Krutko, R.~Refaee, A.~Vijay, and J.~M. Kahn, ``\BIBforeignlanguage{EN}{Ultra-{Low}-{Loss} {Fiber} {Bragg} {Grating} {Mode} {Scrambler} {Design} {Exploiting} {Propagation} {Constant} {Engineering}},'' \emph{\BIBforeignlanguage{EN}{Journal of Lightwave Technology}}, vol.~43, no.~6, pp. 2883--2896, Mar. 2025, publisher: IEEE. [Online]. Available: \url{https://opg.optica.org/jlt/abstract.cfm?uri=jlt-43-6-2883}
\BIBentrySTDinterwordspacing

\bibitem{ho_mode-dependent_2011}
\BIBentryALTinterwordspacing
K.-P. Ho and J.~M. Kahn, ``\BIBforeignlanguage{EN}{Mode-dependent loss and gain: statistics and effect on mode-division multiplexing},'' \emph{\BIBforeignlanguage{EN}{Optics Express}}, vol.~19, no.~17, pp. 16\,612--16\,635, Aug. 2011, publisher: Optica Publishing Group. [Online]. Available: \url{https://opg.optica.org/oe/abstract.cfm?uri=oe-19-17-16612}
\BIBentrySTDinterwordspacing

\bibitem{ho_frequency_2011}
\BIBentryALTinterwordspacing
------, ``Frequency {Diversity} in {Mode}-{Division} {Multiplexing} {Systems},'' \emph{Journal of Lightwave Technology}, vol.~29, no.~24, pp. 3719--3726, Dec. 2011. [Online]. Available: \url{https://ieeexplore.ieee.org/document/6060841}
\BIBentrySTDinterwordspacing

\bibitem{shibahara_long-haul_2020}
\BIBentryALTinterwordspacing
K.~Shibahara, T.~Mizuno, H.~Ono, K.~Nakajima, and Y.~Miyamoto, ``\BIBforeignlanguage{EN}{Long-{Haul} {DMD}-{Unmanaged} 6-{Mode}-{Multiplexed} {Transmission} {Employing} {Cyclic} {Mode}-{Group} {Permutation}},'' in \emph{\BIBforeignlanguage{EN}{Optical {Fiber} {Communication} {Conference} ({OFC}) 2020 (2020), paper {Th3H}.3}}.\hskip 1em plus 0.5em minus 0.4em\relax Optica Publishing Group, Mar. 2020, p. Th3H.3. [Online]. Available: \url{https://opg.optica.org/abstract.cfm?uri=OFC-2020-Th3H.3}
\BIBentrySTDinterwordspacing

\bibitem{di_sciullo_modal_2023}
\BIBentryALTinterwordspacing
G.~Di~Sciullo, M.~Van Den~Hout, G.~Rademacher, R.~S. Luis, B.~J. Puttnam, N.~K. Fontaine, R.~Ryf, H.~Chen, M.~Mazur, D.~T. Neilson, P.~Sillard, F.~Achten, J.~Sakaguchi, C.~Okonkwo, and H.~Furukawa, ``Modal {Dispersion} {Mitigation} in a long-haul 15-{Mode} {Fiber} link through {Mode} {Permutation},'' in \emph{2023 {IEEE} {Photonics} {Society} {Summer} {Topicals} {Meeting} {Series} ({SUM})}, Jul. 2023, pp. 1--2, iSSN: 2376-8614. [Online]. Available: \url{https://ieeexplore.ieee.org/document/10224367}
\BIBentrySTDinterwordspacing

\bibitem{wang_new_2023}
\BIBentryALTinterwordspacing
H.~Wang, X.~Wang, Y.~He, Z.~Yang, Y.~Liu, Q.~Guo, R.~Zhou, X.~Xiao, Z.~Huang, and L.~Zhang, ``\BIBforeignlanguage{EN}{New {Mode}-{Group}-{Permutation} {Strategies} for {MDL} {Reduction} in {Long}-{Haul} {MDM} {Systems}},'' in \emph{\BIBforeignlanguage{EN}{Optical {Fiber} {Communication} {Conference} ({OFC}) 2023 (2023), paper {W3E}.7}}.\hskip 1em plus 0.5em minus 0.4em\relax Optica Publishing Group, Mar. 2023, p. W3E.7. [Online]. Available: \url{https://opg.optica.org/abstract.cfm?uri=OFC-2023-W3E.7}
\BIBentrySTDinterwordspacing

\bibitem{wang_novel_2022}
\BIBentryALTinterwordspacing
Y.~Wang, T.~Gao, Y.~Liu, T.~Xu, W.~Yu, Z.~Yang, Q.~Guo, R.~Zhou, S.~Cao, X.~Xiao, and L.~Zhang, ``\BIBforeignlanguage{EN}{Novel {Mirror}-flipped {Mode} {Permutation} {Technique} for {Long}-haul {Mode}-division {Multiplexing} {Transmissions}},'' in \emph{\BIBforeignlanguage{EN}{Optical {Fiber} {Communication} {Conference} ({OFC}) 2022 (2022), paper {M4B}.5}}.\hskip 1em plus 0.5em minus 0.4em\relax Optica Publishing Group, Mar. 2022, p. M4B.5. [Online]. Available: \url{https://opg.optica.org/abstract.cfm?uri=OFC-2022-M4B.5}
\BIBentrySTDinterwordspacing

\bibitem{xu_modal_2023}
\BIBentryALTinterwordspacing
T.~Xu, Z.~Yang, Y.~Liu, Q.~Guo, R.~Zhou, X.~Xiao, W.~Li, W.~Li, C.~Du, Z.~Huang, and L.~Zhang, ``\BIBforeignlanguage{EN}{Modal {Gain} {Equalization} of {Few}-mode {Erbium}-doped {Fiber} {Amplifiers} {Enabled} by {Mirrored} {Mode} {Exchanges}},'' in \emph{\BIBforeignlanguage{EN}{Optical {Fiber} {Communication} {Conference} ({OFC}) 2023 (2023), paper {M1B}.3}}.\hskip 1em plus 0.5em minus 0.4em\relax Optica Publishing Group, Mar. 2023, p. M1B.3. [Online]. Available: \url{https://opg.optica.org/abstract.cfm?uri=OFC-2023-M1B.3}
\BIBentrySTDinterwordspacing

\bibitem{fan_compact_2022}
\BIBentryALTinterwordspacing
H.~Fan, Z.~Zhang, M.~Cheng, Q.~Yang, M.~Tang, D.~Liu, and L.~Deng, ``\BIBforeignlanguage{EN}{Compact cyclic fiber three-mode converter based on mechanical fiber grating},'' \emph{\BIBforeignlanguage{EN}{Optics Letters}}, vol.~47, no.~17, pp. 4419--4422, Sep. 2022, publisher: Optica Publishing Group. [Online]. Available: \url{https://opg.optica.org/ol/abstract.cfm?uri=ol-47-17-4419}
\BIBentrySTDinterwordspacing

\bibitem{ho_mode_2013}
\BIBentryALTinterwordspacing
K.-P. Ho and J.~M. Kahn, ``\BIBforeignlanguage{en}{Mode {Coupling} and its {Impact} on {Spatially} {Multiplexed} {Systems}},'' in \emph{\BIBforeignlanguage{en}{Optical {Fiber} {Telecommunications}}}.\hskip 1em plus 0.5em minus 0.4em\relax Elsevier, 2013, pp. 491--568. [Online]. Available: \url{https://linkinghub.elsevier.com/retrieve/pii/B9780123969606000110}
\BIBentrySTDinterwordspacing

\bibitem{ho_statistics_2011}
------, ``Statistics of {Group} {Delays} in {Multimode} {Fiber} {With} {Strong} {Mode} {Coupling},'' \emph{Journal of Lightwave Technology}, vol.~29, no.~21, pp. 3119--3128, Nov. 2011, conference Name: Journal of Lightwave Technology.

\bibitem{ho_linear_2014}
\BIBentryALTinterwordspacing
------, ``Linear {Propagation} {Effects} in {Mode}-{Division} {Multiplexing} {Systems},'' \emph{Journal of Lightwave Technology}, vol.~32, no.~4, pp. 614--628, Feb. 2014. [Online]. Available: \url{https://ieeexplore.ieee.org/document/6615972}
\BIBentrySTDinterwordspacing

\bibitem{vijay_modal_2024}
\BIBentryALTinterwordspacing
A.~Vijay, O.~Krutko, R.~Refaee, and J.~M. Kahn, ``Modal {Statistics} in {Mode}-{Division}-{Multiplexed} {Systems} using {Mode} {Scramblers},'' \emph{Journal of Lightwave Technology}, pp. 1--13, 2024, conference Name: Journal of Lightwave Technology. [Online]. Available: \url{https://ieeexplore.ieee.org/document/10685071/?arnumber=10685071}
\BIBentrySTDinterwordspacing

\bibitem{arik_group_2016}
\BIBentryALTinterwordspacing
S.~O. Arık, K.-P. Ho, and J.~M. Kahn, ``Group {Delay} {Management} and {Multiinput} {Multioutput} {Signal} {Processing} in {Mode}-{Division} {Multiplexing} {Systems},'' \emph{Journal of Lightwave Technology}, vol.~34, no.~11, pp. 2867--2880, Jun. 2016, conference Name: Journal of Lightwave Technology. [Online]. Available: \url{https://ieeexplore.ieee.org/document/7409913}
\BIBentrySTDinterwordspacing

\bibitem{jin_mode_2016}
\BIBentryALTinterwordspacing
W.~Jin and K.~S. Chiang, ``\BIBforeignlanguage{EN}{Mode converters based on cascaded long-period waveguide gratings},'' \emph{\BIBforeignlanguage{EN}{Optics Letters}}, vol.~41, no.~13, pp. 3130--3133, Jul. 2016, publisher: Optica Publishing Group. [Online]. Available: \url{https://opg.optica.org/ol/abstract.cfm?uri=ol-41-13-3130}
\BIBentrySTDinterwordspacing

\bibitem{wang_review_2010}
\BIBentryALTinterwordspacing
Y.~Wang, ``Review of long period fiber gratings written by {CO2} laser,'' \emph{Journal of Applied Physics}, vol. 108, no.~8, p. 081101, Oct. 2010, publisher: American Institute of Physics. [Online]. Available: \url{https://aip.scitation.org/doi/10.1063/1.3493111}
\BIBentrySTDinterwordspacing

\bibitem{askarov_long-period_2015}
D.~Askarov and J.~M. Kahn, ``Long-{Period} {Fiber} {Gratings} for {Mode} {Coupling} in {Mode}-{Division}-{Multiplexing} {Systems},'' \emph{Journal of Lightwave Technology}, vol.~33, no.~19, pp. 4032--4038, Oct. 2015, conference Name: Journal of Lightwave Technology.

\bibitem{zhao_broadband_2018}
\BIBentryALTinterwordspacing
Y.~Zhao, H.~Chen, N.~K. Fontaine, J.~Li, R.~Ryf, and Y.~Liu, ``\BIBforeignlanguage{EN}{Broadband and low-loss mode scramblers using {CO}$_{\textrm{2}}$-laser inscribed long-period gratings},'' \emph{\BIBforeignlanguage{EN}{Optics Letters}}, vol.~43, no.~12, pp. 2868--2871, Jun. 2018, publisher: Optica Publishing Group. [Online]. Available: \url{https://opg.optica.org/ol/abstract.cfm?uri=ol-43-12-2868}
\BIBentrySTDinterwordspacing

\bibitem{ma_high-order_2023}
\BIBentryALTinterwordspacing
Y.~Ma, C.~Jiang, Z.~Liu, C.~Mou, and Y.~Liu, ``High-{Order} {OAM} {Mode} {Generator} {Using} {Multi}-{Cascaded} {Long}-{Period} {Fiber} {Gratings},'' \emph{IEEE Photonics Technology Letters}, vol.~35, no.~8, pp. 434--437, Apr. 2023, conference Name: IEEE Photonics Technology Letters. [Online]. Available: \url{https://ieeexplore.ieee.org/document/10064136/?arnumber=10064136}
\BIBentrySTDinterwordspacing

\bibitem{wang_efficient_2024}
\BIBentryALTinterwordspacing
X.~Wang, H.~Guo, Z.~Shi, W.~Chang, Z.~Wang, and Y.-G. Liu, ``Efficient {Mutual} {Conversion} of {High}-{Order} {Core} {Mode} in {Few}-{Mode} {Fiber} {Employing} {Long} {Period} {Fiber} {Gratings},'' \emph{Journal of Lightwave Technology}, vol.~42, no.~7, pp. 2464--2472, Apr. 2024, conference Name: Journal of Lightwave Technology. [Online]. Available: \url{https://ieeexplore.ieee.org/document/10342848/?arnumber=10342848}
\BIBentrySTDinterwordspacing

\bibitem{wang_broadband_2024}
\BIBentryALTinterwordspacing
------, ``\BIBforeignlanguage{en}{Broadband conversion between high-order angular modes based on double-sided exposure long-period fiber grating},'' \emph{\BIBforeignlanguage{en}{Optics Express}}, vol.~32, no.~22, p. 40060, Oct. 2024. [Online]. Available: \url{https://opg.optica.org/abstract.cfm?URI=oe-32-22-40060}
\BIBentrySTDinterwordspacing

\bibitem{zhao_mode_2016}
\BIBentryALTinterwordspacing
Y.~Zhao, Y.~Liu, L.~Zhang, C.~Zhang, J.~Wen, and T.~Wang, ``\BIBforeignlanguage{EN}{Mode converter based on the long-period fiber gratings written in the two-mode fiber},'' \emph{\BIBforeignlanguage{EN}{Optics Express}}, vol.~24, no.~6, pp. 6186--6195, Mar. 2016, publisher: Optica Publishing Group. [Online]. Available: \url{https://opg.optica.org/oe/abstract.cfm?uri=oe-24-6-6186}
\BIBentrySTDinterwordspacing

\bibitem{zhao_mode_2017}
\BIBentryALTinterwordspacing
X.~Zhao, Y.~Liu, Z.~Liu, Y.~Zhao, T.~Wang, L.~Shen, and S.~Chen, ``Mode converter based on the long-period fiber gratings written in the six-mode fiber,'' in \emph{2017 16th {International} {Conference} on {Optical} {Communications} and {Networks} ({ICOCN})}, Aug. 2017, pp. 1--3. [Online]. Available: \url{https://ieeexplore.ieee.org/document/8121432/?arnumber=8121432}
\BIBentrySTDinterwordspacing

\bibitem{zhao_ultra-broadband_2019}
\BIBentryALTinterwordspacing
Y.~Zhao, Z.~Liu, Y.~Liu, C.~Mou, T.~Wang, and Y.~Yang, ``\BIBforeignlanguage{EN}{Ultra-broadband fiber mode converter based on apodized phase-shifted long-period gratings},'' \emph{\BIBforeignlanguage{EN}{Optics Letters}}, vol.~44, no.~24, pp. 5905--5908, Dec. 2019, publisher: Optica Publishing Group. [Online]. Available: \url{https://opg.optica.org/ol/abstract.cfm?uri=ol-44-24-5905}
\BIBentrySTDinterwordspacing

\bibitem{arik_delay_2015}
S.~O. Arık, K.-P. Ho, and J.~M. Kahn, ``Delay {Spread} {Reduction} in {Mode}-{Division} {Multiplexing}: {Mode} {Coupling} {Versus} {Delay} {Compensation},'' \emph{Journal of Lightwave Technology}, vol.~33, no.~21, pp. 4504--4512, Nov. 2015, conference Name: Journal of Lightwave Technology.

\bibitem{vijay_closed-form_2025}
A.~Vijay, N.~Zahedi, O.~Krutko, R.~Refaee, and J.~M. Kahn, ``Closed-{Form} {Statistics} and {Design} of {Mode}-{Division}-{Multiplexing} {Systems} {Employing} {Group}-{Delay} {Compensation} and {Mode} {Permutation},'' 2025.

\bibitem{kogelnik_modal_2012}
\BIBentryALTinterwordspacing
H.~Kogelnik and P.~J. Winzer, ``Modal birefringence in weakly guiding fibers,'' \emph{Journal of Lightwave Technology}, vol.~30, no.~14, pp. 2240--2245, Jul. 2012, conference Name: Journal of Lightwave Technology. [Online]. Available: \url{https://ieeexplore.ieee.org/abstract/document/6193400}
\BIBentrySTDinterwordspacing

\bibitem{fang_low-dmd_2015}
\BIBentryALTinterwordspacing
J.~Fang, A.~Li, and W.~Shieh, ``\BIBforeignlanguage{EN}{Low-{DMD} few-mode fiber with distributed long-period grating},'' \emph{\BIBforeignlanguage{EN}{Optics Letters}}, vol.~40, no.~17, pp. 3937--3940, Sep. 2015, publisher: Optica Publishing Group. [Online]. Available: \url{https://opg.optica.org/ol/abstract.cfm?uri=ol-40-17-3937}
\BIBentrySTDinterwordspacing

\bibitem{lu_full_2010}
\BIBentryALTinterwordspacing
Y.-C. Lu, W.-P. Huang, and S.-S. Jian, ``\BIBforeignlanguage{EN}{Full vector complex coupled mode theory for tilted fiber gratings},'' \emph{\BIBforeignlanguage{EN}{Optics Express}}, vol.~18, no.~2, pp. 713--726, Jan. 2010, publisher: Optica Publishing Group. [Online]. Available: \url{https://opg.optica.org/oe/abstract.cfm?uri=oe-18-2-713}
\BIBentrySTDinterwordspacing

\bibitem{hale_optical_2001}
\BIBentryALTinterwordspacing
A.~Hale, T.~A. Strasser, and P.~S. Westbrook, ``\BIBforeignlanguage{en}{Optical fiber gratings with index matched polymer coatings for cladding mode suppression},'' EP Patent EP1\,146\,357A2, Oct., 2001. [Online]. Available: \url{https://patents.google.com/patent/EP1146357A2/en}
\BIBentrySTDinterwordspacing

\bibitem{meunier_efficient_1991}
\BIBentryALTinterwordspacing
J.~Meunier and S.~Hosain, ``An efficient model for splice loss evaluation in single-mode graded-index fibers,'' \emph{Journal of Lightwave Technology}, vol.~9, no.~11, pp. 1457--1463, Nov. 1991, conference Name: Journal of Lightwave Technology. [Online]. Available: \url{https://ieeexplore.ieee.org/document/97632}
\BIBentrySTDinterwordspacing

\bibitem{liu_reducing_2018}
\BIBentryALTinterwordspacing
H.~Liu, H.~Wen, B.~Huang, R.~A. Correa, P.~Sillard, H.~Chen, Z.~Li, and G.~Li, ``\BIBforeignlanguage{en}{Reducing group delay spread using uniform long-period gratings},'' \emph{\BIBforeignlanguage{en}{Scientific Reports}}, vol.~8, no.~1, p. 3882, Mar. 2018, number: 1 Publisher: Nature Publishing Group. [Online]. Available: \url{https://www.nature.com/articles/s41598-018-21609-1}
\BIBentrySTDinterwordspacing

\bibitem{sillard_low-differential-mode-group-delay_2016}
\BIBentryALTinterwordspacing
P.~Sillard, D.~Molin, M.~Bigot-Astruc, K.~De~Jongh, F.~Achten, A.~M. Velázquez-Benítez, R.~Amezcua-Correa, and C.~M. Okonkwo, ``Low-{Differential}-{Mode}-{Group}-{Delay} 9-{LP}-{Mode} {Fiber},'' \emph{Journal of Lightwave Technology}, vol.~34, no.~2, pp. 425--430, Jan. 2016, conference Name: Journal of Lightwave Technology. [Online]. Available: \url{https://ieeexplore.ieee.org/abstract/document/7174947}
\BIBentrySTDinterwordspacing

\bibitem{choutagunta_designing_2021}
K.~Choutagunta and J.~M. Kahn, ``Designing {High}-{Performance} {Multimode} {Fibers} {Using} {Refractive} {Index} {Optimization},'' \emph{Journal of Lightwave Technology}, vol.~39, no.~1, pp. 233--242, Jan. 2021, conference Name: Journal of Lightwave Technology.

\bibitem{ostling_broadband_1996}
\BIBentryALTinterwordspacing
D.~Östling and H.~E. Engan, ``\BIBforeignlanguage{EN}{Broadband spatial mode conversion by chirped fiber bending},'' \emph{\BIBforeignlanguage{EN}{Optics Letters}}, vol.~21, no.~3, pp. 192--194, Feb. 1996, publisher: Optica Publishing Group. [Online]. Available: \url{https://opg.optica.org/ol/abstract.cfm?uri=ol-21-3-192}
\BIBentrySTDinterwordspacing

\bibitem{scarnera_splice_2024}
\BIBentryALTinterwordspacing
V.~Scarnera, C.~A. Codemard, M.~Durkin, and M.~N. Zervas, ``Splice optimisation between dissimilar fibres in the presence of dopant diffusion,'' in \emph{Fiber {Lasers} {XXI}: {Technology} and {Systems}}, vol. 12865.\hskip 1em plus 0.5em minus 0.4em\relax SPIE, Mar. 2024, pp. 197--201. [Online]. Available: \url{https://www.spiedigitallibrary.org/conference-proceedings-of-spie/12865/1286515/Splice-optimisation-between-dissimilar-fibres-in-the-presence-of-dopant/10.1117/12.3003118.full}
\BIBentrySTDinterwordspacing

\end{thebibliography}

\begin{IEEEbiographynophoto}{Oleksiy Krutko}
received the B.S. degree in electrical engineering from the University of Texas at Austin, Austin, TX, USA, in 2020. He is currently working toward the Ph.D. degree from Stanford University, Stanford, CA, USA. His research interests include optical fiber communications and photonic devices.
\end{IEEEbiographynophoto}

\begin{IEEEbiographynophoto}{Rebecca Refaee}
received a B.S. degree in Mathematics and an M.S. degree in Electrical Engineering from Stanford University, Stanford, CA, USA in 2024. 
She is currently working towards a Ph.D. degree in Electrical Engineering at Stanford University. Her current research interests include optical communications and mode-division multiplexing.
\end{IEEEbiographynophoto}

\begin{IEEEbiographynophoto}{Anirudh Vijay}
received the B.Tech. and M.Tech. degrees in Electrical Engineering from the Indian Institute of Technology Madras, Chennai, Tamil Nadu, India, in 2019.
He is working towards the Ph.D. degree in Electrical Engineering from Stanford University, Stanford, CA, USA. 
His current research interests include optical communications, mode-division multiplexing, and data-center applications.
\end{IEEEbiographynophoto}

\begin{IEEEbiographynophoto}{Nika Zahedi}
is currently working towards the B.S. and M.S. degrees in electrical engineering at Stanford University. Her current research interests include mode-division multiplexing, signal processing, and optimization techniques.
\end{IEEEbiographynophoto}

\begin{IEEEbiographynophoto}{Joseph M. Kahn}
(F’00) received A.B., M.A. and Ph.D. degrees in Physics from the University of California, Berkeley in 1981, 1983 and 1986. In 1987-1990, Kahn was at AT\&T Bell Laboratories. In 1989, he demonstrated the first successful synchronous (i.e., coherent) detection using semiconductor lasers, achieving record receiver sensitivity. In 1990-2003, Kahn was on the Electrical Engineering and Computer Sciences faculty at Berkeley. He demonstrated coherent detection of QPSK in 1992. In 1999, D. S. Shiu and Kahn published the first work on probabilistic shaping for optical communications. In the 1990s and early 2000s, Kahn and collaborators performed seminal work on indoor and outdoor free-space optical communications and multi-input multi-output wireless communications. In 2000, Kahn and K. P. Ho founded StrataLight Communications, whose 40 Gb/s-per-wavelength long-haul fiber transmission systems were deployed widely by AT\&T, Deutsche Telekom, and other carriers. In 2002, Ho and Kahn applied to patent the first electronic compensation of fiber Kerr nonlinearity. StrataLight was acquired by Opnext in 2009. In 2003, Kahn became a Professor of Electrical Engineering in the E. L. Ginzton Laboratory at Stanford University. Kahn and collaborators have extensively studied rate-adaptive coding and modulation, as well as digital signal processing for mitigating linear and nonlinear impairments in coherent systems. In 2008, E. Ip and Kahn (and G. Li independently) invented simplified digital backpropagation for compensating fiber Kerr nonlinearity and dispersion. Since 2004, Kahn and collaborators have studied propagation, modal statistics, spatial multiplexing and imaging in multi-mode fibers, elucidating principal modes and demonstrating transmission beyond the traditional bandwidth-distance limit in 2005, deriving the statistics of coupled modal group delays and gains in 2011, and deriving resolution limits for imaging in 2013. Kahn’s current research addresses optical frequency comb generators, coherent data center links, rate-adaptive access networks, fiber Kerr nonlinearity mitigation, ultra-long-haul submarine links, and optimal free-space transmission through atmospheric turbulence. Kahn received the National Science Foundation Presidential Young Investigator Award in 1991. In 2000, he became a Fellow of the IEEE.
\end{IEEEbiographynophoto}

\end{document}